\documentclass[prd,eqsecnum,twocolumn,amsfonts,amssymb]{revtex4}

\usepackage{graphicx}

\usepackage{bm}

\newcommand{\beq}{\begin{equation}}
\newcommand{\eeq}{\end{equation}}
\newcommand{\beqs}{\begin{eqnarray}}
\newcommand{\eeqs}{\end{eqnarray}}

\begin{document}

\title{Scheme-Independent Calculations of Properties at a Conformal Infrared 
Fixed Point in Gauge Theories with Multiple Fermion Representations}

\author{Thomas A. Ryttov$^a$ and Robert Shrock$^b$}

\affiliation{(a) \ CP$^3$-Origins, University of Southern Denmark, \\
Campusvej 55, Odense, Denmark}

\affiliation{(b) \ C. N. Yang Institute for Theoretical Physics and 
Department of Physics and Astronomy, \\
Stony Brook University, Stony Brook, NY 11794, USA }

\begin{abstract}

  In previous work we have presented scheme-independent calculations of
  physical properties of operators at a conformally invariant infrared fixed
  point in an asymptotically free gauge theory with gauge group $G$ and $N_f$
  fermions in a representation $R$ of $G$.  Here we generalize this analysis to
  the case of fermions in multiple representations, focusing on the case of two
  different representations. Our results include the calculation of the
  anomalous dimensions of gauge-invariant fermion bilinear operators, and the
  derivative of the beta function, evaluated at the infrared fixed point. We
  illustrate our results in an SU($N_c$) gauge theory with $N_F$ fermions in
  the fundamental representation and $N_{Adj}$ fermions in the adjoint
  representation.

\end{abstract}

\maketitle


\section{Introduction}
\label{intro_section}

In this paper we shall consider a vectorial, asymptotically free gauge theory
(in four spacetime dimensions, at zero temperature) with gauge group $G$ with
massless fermions transforming according to multiple different representations
of $G$, which has an exact infrared (IR) fixed point (IRFP) of the
renormalization group \cite{fm}. For technical simplicity, we will restrict
ourselves to two different representations. We thus take the theory to contain
$N_f$ copies (flavors) of Dirac fermions, denoted $f$, in the
representation $R$ of $G$, and $N_{f'}$ copies of fermions, denoted $f'$, in a
different representation $R'$ of $G$.  In the case in which $f'$ transforms
according to a self-conjugate representation, the number $N_{f'}$ refers
equivalently to a theory with $N_{f'}$ Dirac fermions or $2N_{f'}$ Majorana
fermions and hence in this case $N_{f'}$ may take on half-integral as well as
integral values.  One motivation for such theories is a possible direction for
ultraviolet completions of the Standard Model (e.g., \cite{rs09,bv}). In
\cite{bv} we studied the infrared evolution and phase structure of this type of
theory.  Here we go beyond Refs. \cite{rs09,bv} in presenting
(scheme-independent) calculations of anomalous dimensions of gauge-invariant
operators.

We denote the running gauge coupling as $g=g(\mu)$, where $\mu$ is the
Euclidean energy/momentum scale at which this coupling is measured.  We define
$\alpha(\mu) = g(\mu)^2/(4\pi)$.  Since the theory is asymptotically free, its
properties can be computed reliably in the deep ultraviolet (UV) region at
large $\mu$, where the coupling approaches zero.  The dependence of
$\alpha(\mu)$ on $\mu$ is described by the renormalization-group (RG) beta
function, $\beta = d\alpha(\mu)/dt$, where $dt=d\ln\mu$ (the argument $\mu$
will often be suppressed in the notation). We will consider a theory in which
the fermion content is such that the RG flow from the UV to the IR ends in an
exact IR fixed point, as determined by the zero in the beta function nearest to
the origin for physical coupling, denoted $\alpha_{IR}$. Since $\beta=0$ at
$\alpha=\alpha_{IR}$, the resultant theory in this IR limit is scale-invariant,
and is deduced also to be conformally invariant \cite{scalecon}.

The properties of the resultant conformal field theory at this IRFP are of
considerable importance. Physical quantities defined at the IRFP obviously
cannot depend on the scheme used for the regularization and renormalization of
the theory.  In conventional computations of these quantities, one first writes
them as series expansions in powers of the coupling, and then evaluates these
series expansions with $\alpha$ set equal to $\alpha_{IR}$, calculated to a
given loop order.  These calculations have been performed for anomalous
dimensions of gauge-invariant fermion bilinears in a theory with a single
fermion representation up to four-loop level \cite{bvh}-\cite{bc} and to
five-loop level \cite{flir}.  However, as is well known, these conventional
(finite-order) series expansions are scheme-dependent beyond the leading terms.
Indeed, this is a generic property of higher-order calculations in quantum
field theory, such as computations in quantum chromodynamics (QCD) used to
compare with data from the Fermilab Tevatron and CERN Large Hadron Collider
(LHC).

There is thus strong motivation to calculate and analyze series expansions for
physical properties at the IRFP which are scheme-independent at each
finite order. The fact that makes this possible is simple but powerful.  To
review this, we first specialize to a theory with $N_f$ fermions in a single
representation, $R$, of the gauge group $G$.  The constraint of asymptotic
freedom means that $N_f$ must be less than a certain upper ($u$) bound, denoted
$N_{f,u}$. Here and below, we will often formally generalize the number(s) of
fermions in one or multiple representations from non-negative integers to
non-negative real numbers, with the understanding that for a physical quantity
one restricts to integral values. Furthermore, as noted above, if an $f'$
fermion transforms according to a self-conjugate representation, then 
the number $N_{f'}$ refers equivalently to a theory with $N_{f'}$ Dirac
fermions or $2N_{f'}$ Majorana fermions, so that in this case, $N_{f'}$ may
take on half-integral physical values.  As $N_f$ approaches $N_{f,u}$
from below, the value of the IRFP, $\alpha_{IR}$, approaches zero. 
This means that one can re-express series expansions for physical
quantities at this IRFP in powers of the manifestly scheme-independent variable
\cite{bz,gkgg}
\beq
\Delta_f = N_{f,u}-N_f \ . 
\label{deltaf}
\eeq
In recent work, for theories with $N_f$ fermions in a single representation of
the gauge group $G$, we have calculated scheme-independent series
expansions for the anomalous dimensions of gauge-invariant fermion bilinears
and the derivative $d\beta/d\alpha$, both evaluated at the IRFP, to the
respective orders $O(\Delta_f^4)$ and $O(\Delta_f^5)$
\cite{gtr}-\cite{bpcgt}. These are the highest orders to which these quantities
have been calculated. We gave explicit expressions for the case $G={\rm
  SU}(N_c)$ and $R$ equal to the fundamental, adjoint, and rank-2 symmetric and
antisymmetric tensor representations, and for other Lie groups, including
orthogonal, symplectic, and exceptional groups.

In this paper we shall generalize our previous scheme-independent series
calculations of physical quantities at an IRFP from the case of an
asymptotically free gauge theory with $N_f$ fermions in a single representation
of the gauge group $G$ to the case of fermions in multiple different
representations.  Specifically, we consider a theory with $N_f$ fermions in a
representation $R$ of $G$ and $N_{f'}$ fermions in a different representation,
$R'$, of $G$.  We present scheme-independent calculations of the anomalous
dimensions of gauge-invariant fermion bilinear operators to cubic order in the
respective expansion variable ($\Delta_f$ in Eq. (\ref{deltaf}) for 
$\bar f f $ and $\Delta_{f'}$ in Eq. (\ref{deltaf2}) for $\bar f' f'$) and 
to quartic order
in $\Delta_f$ and $\Delta_{f'}$ for the derivative of the beta function,
evaluated at the infrared fixed point.

The condition of asymptotic freedom requires that the value of a certain linear
combination of $N_{f'}$ and $N_f$ must be less than an upper bound given below
by Eq. (\ref{asymptotic_freedom}). For a fixed $N_{f'}$, this implies an upper
bound denoted as $N_f < N_{f,u}$, and for a fixed $N_f$, this implies the upper
bound $N_{f'} < N_{f',u}$ given respectively in Eqs. (\ref{Nfu}) and
(\ref{Nf2u}) below. For fixed $N_{f'}$, as $N_f$ approaches $N_{f,u}$ from
below, $\alpha_{IR}$ approaches zero. Therefore, one can rewrite the series
expansions for physical quantities as power series in the variable
$\Delta_f$. The coefficients in these series expansions depend on $N_{f'}$.
If $\Delta_f$ is small, the value of $\alpha_{IR}$ is also small, so that the
resultant IR theory may be inferred to be in a (deconfined) non-Abelian Coulomb
phase (NACP), often called the conformal window.  Strong evidence for this in
the single-representation case comes from fully nonperturbative lattice
simulations \cite{lgtreviews,simons,lnncomment}. In the same way, for fixed
$N_f$, one can rewrite the series expansions for physical quantities as power
series in the variable
\beq
\Delta_{f'} = N_{f',u} - N_{f'} \ . 
\label{deltaf2}
\eeq

For a general operator ${\cal O}$, we denote the full scaling dimension as
$D_{\cal O}$ and its free-field value as $D_{{\cal O},free}$.  The anomalous
dimension of this operator, denoted $\gamma_{\cal O}$, is defined via the
relation \cite{gammaconvention}
\beq
D_{\cal O} = D_{{\cal O},free} - \gamma_{\cal O} \ . 
\label{anomdim}
\eeq
Let us denote the fermions of type $f$ as $\psi_i$, $i=1,...,N_f$ and the
fermions of type $f'$ as $\chi_j$, $j=1,...,N_{f'}$.  We shall calculate
scheme-independent series expansions for the anomalous dimensions, denoted
$\gamma_{\bar\psi\psi,IR}$ and $\gamma_{\bar\chi\chi,IR}$ of the respective
(gauge-invariant) fermion bilinears
\beq
\bar\psi\psi = \sum_{j=1}^{N_f} \bar\psi_j \psi_j
\label{psibarpsi}
\eeq
and 
\beq
\bar\chi\chi = \sum_{j=1}^{N_{f'}} \bar\chi_j \chi_j \ .
\label{chibarchi}
\eeq
The anomalous dimension of $\bar\psi\psi$ is the same as that of the
(gauge-invariant) bilinear $\sum_{j,k=1}^{N_f} \bar\psi_j {\cal T}_a \psi_k$,
where ${\cal T}_a$ is a generator of the Lie algebra of SU($N_f$)
\cite{gracey_gammatensor}, and we shall use the symbol
$\gamma_{\bar\psi\psi,IR}$ to refer to both.  An analogous comment applies to
$\gamma_{\bar\chi\chi,IR}$. We write the scheme-independent series expansions
of $\gamma_{\bar f f,IR}$ as 
\beq
\gamma_{\bar f f ,IR} = \sum_{j=1}^\infty \kappa^{(f)}_j \, \Delta_f^j
\label{gamma_ir_Deltaseries}
\eeq
and
\beq
\gamma_{\bar f' f' ,IR} = \sum_{j=1}^\infty \kappa^{(f')}_j \, \Delta_{f'}^j
\label{gamma_ir_Delta2series}
\eeq

We shall illustrate our general results in an SU($N_c$) gauge theory with $N_F$
fermions of type $f$ in the fundamental ($F$) representation and $N_{Adj}$
fermions of type $f'$ in the adjoint ($Adj$) representation.
For this theory we will also use an explicit notation with
coefficients $\kappa^{(f)} = \kappa^{(F)}$ and $\kappa^{(f')}=\kappa^{(Adj)}$. 

We shall calculate two equivalent scheme-independent series expansions of the 
derivative $\beta'_{IR}$. With $N_{f'}$ fixed, and $N_f$ variable, one may
write the series as an expansion in powers of $\Delta_f$:
\beq
\beta'_{IR} = \sum_{j=2}^\infty d_j \, \Delta_f^j \ .  
\label{betaprime_ir_Deltaseries}
\eeq
Alternately, one may take $N_f$ to be
fixed and write $\beta'_{IR}$ as a series expansion in powers of $\Delta_{f'}$,
as
\beq \beta'_{IR} = \sum_{j=2}^\infty \tilde d_j \, \Delta_{f'}^j \ .
\label{betaprime_ir_Deltaseries2}
\eeq
Note that $d_1 = \tilde d_1 = 0$ for all $G$ and fermion representations. 

This paper is organized as follows. In Section \ref{methods_section} we discuss
the methodology for our calculations. In Sections \ref{kappa_section} and
\ref{betaprime_section} we present our new results for scheme-independent
expansions of the anomalous dimensions of fermion bilinears and
$d\beta/d\alpha$, both evaluated at the infrared fixed point.  We discuss the
special cases of the anomalous dimension and $\beta'_{IR}$ results for an
illustrative theory with gauge group SU($N_c$) containing fermions in the
fundamental and adjoint representations in Sections \ref{kappa_fadj_section}
and \ref{betaprime_fadj_section}, respectively.  Our conclusions are given in
Section \ref{conclusion_section}, and some relevant group-theoretic results are
reviewed in Appendix \ref{groupinvariants}.


\section{Calculational Methods} 
\label{methods_section}

\subsection{Beta Function and Series Expansions for Physical Quantities} 

In this section we discuss some background and the calculational methods that
are relevant for our present work.  The series expansion of $\beta$ in powers
of the squared gauge coupling is
\beq
\beta = -2\alpha \sum_{\ell=1}^\infty b_\ell \, a^\ell \ , 
\label{beta}
\eeq
where $a=g^2/(16\pi^2) = \alpha/(4\pi)$ and $b_\ell$ is the $\ell$-loop
coefficient. With an overall minus sign extracted, as in Eq. (\ref{beta}), the
condition of asymptotic freedom is that $b_1 > 0$.  The one-loop coefficient,
$b_1$, is independent of the scheme used for regularization and
renormalization. Mass-independent schemes include minimal subtraction \cite{ms}
and modified minimal subtraction, denoted $\overline{\rm MS}$ \cite{msbar}.
For mass-independent schemes, the two-loop coefficient, $b_2$, is also
independent of the specific scheme used \cite{gross75}.  For a theory with a
general gauge group $G$ and $N_f$ fermions in a single representation, $R$, the
coefficients $b_1$ and $b_2$ were calculated in \cite{b1} and \cite{b2}, while
$b_3$, $b_4$, and $b_5$ were calculated in the commonly used $\overline{\rm
  MS}$ scheme in \cite{b3}, \cite{b4}, and \cite{b5}, respectively 
(see also \cite{b5su3}). For the analysis of a theory with fermions in multiple
different representations, one needs generalizations of these results.  These
are straightforward to derive in the case of $b_1$ and $b_2$, but new
calculations are required for higher-loop coefficients.  These have recently
been performed in \cite{zoller} (again in the $\overline{\rm MS}$ scheme) up to
four-loop order, and we use the results of Ref. \cite{zoller} here. 

The expansion of the anomalous dimension of the fermion bilinear
$\gamma_{\bar\psi\psi}$ in powers of the squared gauge coupling is 
\beq
\gamma_{\bar\psi\psi} = \sum_{\ell=1}^\infty c^{(f)}_\ell a^\ell \ , 
\label{gamma}
\eeq
where $c^{(f)}_\ell$ is the $\ell$-loop coefficient. The analogous expansion
applies for $\gamma_{\bar\chi\chi}$ with the replacement $c^{(f)}_\ell \to
c^{(f')}_\ell$. The one-loop coefficient $c^{(f)}_1$ is scheme-independent,
while the $c^{(f)}_\ell$ with $\ell \ge 2$ are scheme-dependent, and similarly
with the $c^{(f')}_\ell$.  For a general gauge group $G$ and $N_f$ fermions in
a single representation $R$ of $G$, the $c^{(f)}_\ell$ have been calculated up
to loop order $\ell=4$ in \cite{c4} and $\ell=5$ in \cite{c5}. For the case of
multiple fermion representations, the anomalous dimension coefficients for the
fermion bilinears have been calculated up to four-loop order in
\cite{chetzol}. We use the results of \cite{chetzol} up to three-loop order
here.

Concerning scheme-independent series expansions, the calculation of the
coefficient $\kappa^{(f)}_j$ in Eq. (\ref{gamma_ir_Deltaseries}) requires, as
inputs, the values of the $b_\ell$ for $1 \le \ell \le j+1$ and the
$c^{(f)}_\ell$ for $1 \le \ell \le j$, and similarly for $\kappa^{(f')}_j$,
with the replacement $c^{(f)}_\ell \to c^{(f')}_\ell$. The calculation of the
coefficients $d_j$ and $\tilde d_j$ in Eqs. (\ref{betaprime_ir_Deltaseries})
and (\ref{betaprime_ir_Deltaseries2}) requires, as inputs, the values of the
$b_\ell$ for $1 \le \ell \le j$.  

Thus, using the calculation of the beta
function for multiple fermion representation to four-loop order in
\cite{zoller}, together with the calculation of the anomalous dimensions of the
fermion bilinears in \cite{chetzol} up to three-loop order, 
we can calculate $\gamma_{\bar\psi\psi,IR}$ to order $O(\Delta_f^3)$ and
$\gamma_{\bar\chi\chi,IR}$ to $O(\Delta_{f'}^3)$ for the case of multiple
fermion representations.  (Note that we cannot make use of the four-loop
calculation of the anomalous dimensions of fermion bilinears in \cite{chetzol}
to compute $\gamma_{\bar\psi\psi,IR}$ to order $O(\Delta_f^4)$ and
$\gamma_{\bar\chi\chi,IR}$ to $O(\Delta_{f'}^4)$, because this would require,
as an input, the five-loop coefficient $b_5$ in the beta function for this case
of multiple fermion representations, and, to our knowledge, this has not been
calculated.) 

Similarly, using the four-loop beta function from \cite{zoller}, we can
calculate the $d_j$ and $\tilde d_j$ for $\beta'_{IR}$ to order $j=4$. We
denote the truncation of these series to maximal power $j=p$ as
$\gamma_{\bar\psi\psi,IR,\Delta_f^p}$, $\gamma_{\bar\chi\chi,IR,\Delta_f^p}$,
$\beta'_{IR,\Delta_f^p}$, and $\beta'_{IR,\Delta_{f'}^p}$, respectively.
Although we use these coefficients as calculated in the $\overline{\rm MS}$
scheme below, we emphasize that our results are scheme-independent, so the
specific scheme used for their calculation does not matter.  An explicit
illustration of this using several schemes is given in \cite{cgs}. We refer the
reader to our previous work for detailed discussions of the procedure for
calculating the coefficients $\kappa_j$ and $d_j$ in the case of a theory with
$N_f$ fermions in a single representation of $G$.

Our procedure for calculating scheme-independent series expansions requires
that the IRFP be exact, and hence we restrict our consideration to the
non-Abelian Coulomb phase, where this condition is satisfied. For sufficiently
smaller values of $N_f$ and/or $N_{f'}$, there is spontaneous chiral symmetry
breaking (S$\chi$SB), giving rise to dynamical masses for the $f$ and/or $f'$
fermions \cite{phasenote}. Most-attractive channel arguments suggest that as
$N_f$ and/or $N_{f'}$ decrease(s) and $\alpha_{IR}$ increases, the fermion with
the largest value of $C_f$ would be the first to form bilinear fermion
condensates and hence obtain dynamical masses and be integrated out of the
low-energy effective field theory (EFT).  Assuming that this happens and,
say, the $f'$ fermions condense out, then one would proceed to examine the
resultant EFT with the remaining massless $f$ fermions to determine the
further evolution of this theory into the infrared. The details of the
construction of the EFT will not be relevant here, since we restrict our
analysis to the (chirally symmetric) non-Abelian Coulomb phase. 


\subsection{Relevant Range of $(N_f,N_{f'})$} 

Since we require that the theory should be asymptotically free and since our
scheme-independent calculational method requires an exact IR fixed point, which
is satisfied in the non-Abelian Coulomb phase, a first step is to discuss the
corresponding values of the pair $(N_f,N_{f'})$ that satisfy these
conditions. We denote this set of values, or more generally, the region in the
first quadrant of the ${\mathbb R}^2$ plane defined by the generalization of
$(N_f,N_{f'})$ from non-negative integers (or half-integers in the case of a
Majorana fermion in a self-conjugate representation) to non-negative real
numbers, where the theory has an IRFP in the non-Abelian Coulomb phase as the
region ${\cal R}_{NACP}$.  We next discuss the boundaries of this region.

For a specified gauge group $G$ and fermion representations $R$ and $R'$, the 
numbers $N_f$ and $N_{f'}$ are bounded above by the asymptotic freedom (AF)
condition that $b_1 > 0$. This condition is expressed as the inequality on the
linear combination 
\beq
N_fT_f + N_{f'}T_{f'} < \frac{11C_A}{4} \ , 
\label{asymptotic_freedom}
\eeq
where $C_A$ and $T_f$ are group invariants defined in Appendix
\ref{groupinvariants}. Thus, for fixed $N_{f'}$, the AF property implies that 
$N_f$ is bounded above as $N_f < N_{f,u}$, where
\beq
N_{f,u} = \frac{11C_A - 4N_{f'}T_{f'}}{4T_f} \ ,
\label{Nfu}
\eeq
and similarly, for fixed $N_f$, the AF condition implies that $N_{f'}$ is 
bounded above as $N_{f'} < N_{f',u}$, where 
\beq
N_{f',u} = \frac{11C_A - 4N_fT_f}{4T_{f'}} \ . 
\label{Nf2u}
\eeq

The upper boundary of this asymptotically free region, which is also the upper
boundary of the region ${\cal R}_{NACP}$, in $N_f$ and
$N_{f'}$ is the locus of solutions to the condition $b_1=0$. This is a finite
segment of the line $N_fT_f + N_{f'}T_{f'} = 11C_A/4$.  We may picture the
first quadrant in the ${\mathbb R}^2$ space defined by non-negative 
$(N_f,N_{f'})$ to be such that $N_f$ is the
horizontal axis and $N_{f'}$ is the vertical axis. Then the line segment
bounding the asymptotically free region is an oblique line segment running from
the upper left to the lower right, with slope
\beq
\frac{\partial N_{f'}}{\partial N_f}\bigg |_{b_1=0} = -\frac{T_f}{T_{f'}} \ . 
\label{slope_b1zeroline}
\eeq
This line segment intersects the horizontal axis at the point
$(N_f,N_{f'})=(11C_A/(4T_f),0)$ and the vertical axis at the point
$(N_f,N_{f'})=(0,11C_A/(4T_{f'}))$. Without loss of generality, we take $f$ to
be the (nonsinglet) fermion representation of smaller dimension.  The
respective scheme-independent expansions in powers of $\Delta_f$ and
$\Delta_{f'}$ amount to moving into the interior of the non-Abelian Coulomb
phase from the upper boundary line horizontally (moving leftward) and
vertically (moving downward).

In our earlier work on theories with $N_f$ fermions in a single fermion
representation of the gauge group, we denoted the lower boundary of the NACP as
$N_{f,cr}$.  In that case, we assumed that $N_f$ was in the NACP interval
$I_{NACP}: \ N_{f,cr} < N_f < N_{f,u}$.  Here the generalization of this is the
set of physical values of $N_f$ and $N_{f'}$ in the region ${\cal
  R}_{NACP}$. Even in the case of a single fermion representation, the value of
$N_{f,cr}$ is not known precisely. This question of the value of $N_{f,cr}$ for
various specific theories has been investigated in a number of lattice studies
\cite{lgtreviews,simons}, which continue at present.  As noted above, we have
previously presented approximate analytic results relevant for this study in
\cite{rs09,bv} Corresponding lattice studies could be carried out for theories
with multiple different fermion representations to study properties of the
respective theories. An example is a recent lattice study of an SU(4) gauge
theory with $N_f=2$ Dirac fermions in the fundamental representation and
$N_{f'}=2$ Dirac fermions in the (self-conjugate) antisymmetric rank-2 tensor
representation \cite{su4lgt1,su4lgt2}, which finds that the (zero-temperature)
theory is in the phase with chiral symmetry breaking for both types of
fermions. Since our results are restricted to an exact infrared fixed point in
the (conformally invariant) non-Abelian Coulomb phase, they are not directly
applicable to this theory. 

For the present study, with the axes of the first-quadrant quarter plane in
$(N_f,N_{f'}) \in {\mathbb R}^2$ as defined above, the upper boundary of the
NACP is the line segment resulting from the $b_1=0$ condition.  The analogue of
the lower boundary of the NACP at $N_{f,cr}$ for the present study with two
fermion representations is a line segment or nonlinear curve displaced in the
direction to the lower left relative to the oblique $b_1=0$ line, so that the
resultant NACP forms a region in which physical values of $N_f$ and $N_{f'}$
define possible IR theories.  This lower boundary of the NACP intersects the
horizontal axis at the point $(N_f,N_{f'})=(N_{f,cr},0)$ and intersects the
vertical axis at the point $(N_f,N_{f'})=(0,N_{f',cr})$.  Although this lower
boundary of the NACP is not known, one can get a rough idea of where it lies by
generalizing the analysis that we gave in our previous work for theories with a
single fermion representation \cite{gsi,dex,dexl}. This analysis was based on
the observation that the two-loop beta function has an IR zero if $N_f$ is
sufficiently large that $b_2$ is negative (with $b_1 > 0$).  In this case of a
single fermion representation, for small $N_f$, $b_2$ is positive, and turns
negative when $N_f$ exceeds a certain lower ($\ell$) value $N_{f,\ell} <
N_{f,u}$ where $b_2=0$, namely
\beq
N_{f,\ell} = \frac{17C_A^2}{2T_f(5C_A+3C_f)}  \quad ({\rm for} \ N_{f'}=0).
\label{nell}
\eeq
Thus, in this single-representation case, if and only if $N_f$ lies in an
interval that we have denoted previously as $I_{IRZ}$, the two-loop beta
function has an IR zero (IRZ). This interval $I_{IRZ}$ is
\beq
I_{IRZ}: \quad N_{f,\ell} < N_f < N_{f,u} \quad  ({\rm for} \ N_{f'}=0). 
\label{interval}
\eeq
Although $N_{f,\ell}$ is not, in general, equal to $N_{f,cr}$, it is moderately
close to the latter in theories that have been studied. As an example, in the
case of an SU($N_c$) gauge theory with
$N_f$ fermions in the fundamental ($F$) representation,
\beq
{\rm SU}(N_c), \ R=F: \quad N_{f,\ell} = \frac{34N_c^3}{13N_c^2-3} \ . 
\label{nell_sun}
\eeq
In the intensively studied case $N_c=3$ theory, $N_\ell = 153/19 \simeq 8.05$.
This is close to the estimates of $N_{f,cr}$ for this theory from our previous
studies and from a number of lattice simulations
\cite{gsi,dexl,lgtreviews,simons}.

In our present asymptotically free theory with two fermion representations, 
the two-loop beta function has an IR zero if and only if $b_2 < 0$, which is
the inequality 
\beqs
&&N_fT_f(5C_A+3C_f) + N_{f'}T_{f'}(5C_A+3C_{f'}) > \frac{17C_A^2}{2} \ . \cr\cr
&&
\label{b2_negative}
\eeqs
This IR zero of the two-loop ($2\ell$) beta function occurs at 
$\alpha=\alpha_{IR,2\ell}$, where 
\begin{widetext}
\beq
\alpha_{IR,2\ell}=-\frac{4\pi b_1}{b_2} 
=\frac{2\pi\Big [11C_A - 4(N_fT_f + N_{f'}T_{f'})\Big ] }
{\Big [2N_fT_f(5C_A+3C_f) + 2N_{f'}T_{f'}(5C_A+3C_{f'}) - 17C_A^2\Big ]} \ . 
\label{alfir_2loop_ff2}
\eeq
\end{widetext}
We thus define the two-dimensional region in the first quadrant of the
${\mathbb R}^2$ plane defined by non-negative real values of $(N_f,N_{f'})$
where the theory is asymptotically free and the two-loop beta function has an
IR zero as the region ${\cal R}_{IRZ}$, given by the conditions
(\ref{asymptotic_freedom}) and (\ref{b2_negative}).  The upper boundary of
${\cal R}_{IRZ}$ is the same as the upper boundary of ${\cal R}_{NACP}$, while
the lower boundary of ${\cal R}_{IRZ}$ can provide a rough guide to the lower
boundary of ${\cal R}_{NACP}$ and has the advantage that it is exactly
calculable. This lower boundary of the region ${\cal R}_{IRZ}$ is given by the
solution of the condition that $b_2=0$ in the first quadrant of the ${\mathbb
  R}^2$ plane. This condition is obtained from Eq. (\ref{b2_negative}) by
replacing the inequality by an equality. The corresponding line defining the
lower boundary of ${\cal R}_{IRZ}$ has the slope
\beq
\frac{\partial N_{f'}}{\partial N_f}{}\bigg |_{b_2=0} = 
-\frac{T_f(5C_A+3C_f)}{T_{f'}(5C_A+3C_{f'})} \ .
\label{slope_b2zeroline}
\eeq
This lower boundary of the region ${\cal R}_{IRZ}$ crosses the horizontal 
axis in the $(N_f,N_{f'})$ space at the point $(N_{f,\ell},0)$, where 
$N_{f,\ell}$ was given above in Eq. (\ref{nell}), and it crosses
the vertical axis at the corresponding value $(0,N_{f',\ell})$, where 
\beq
N_{f',\ell} = \frac{17C_A^2}{2T_{f'}(5C_A+3C_{f'})} \ . 
\label{nf2ell}
\eeq
As noted, the lower boundary of this ${\cal R}_{IRZ}$ region provides a rough
guide to the actual lower boundary of the NACP region ${\cal R}_{NACP}$. The
determination of the true lower boundary of ${\cal R}_{NACP}$ would require
a fully nonperturbative analysis, e.g., via lattice simulations. 

Although our calculational methods require the IRFP to be exact and hence,
strictly speaking, apply only in the non-Abelian Coulomb phase, they could also
be useful for the investigation of quasi-conformal gauge theories. In turn, the
latter have been of interest as possible ultraviolet completions of the
Standard Model. Specifically, (a) if the transition from the lower part of the
non-Abelian Coulomb phase to the quasi-conformal regime in the variables
$(N_f,N_{f'})$ is continuous, and (b) if our series calculations are
sufficiently accurate in this region, our results for
$\gamma_{\bar\psi\psi,IR}$, $\gamma_{\bar\chi\chi,IR}$, and $\beta'_{IR}$ could
provide approximate estimates for the values of these quantities in the
quasi-conformal regime just below the lower boundary with the NACP.


\subsection{Example with Fermions in the Fundamental and Adjoint
  Representations} 

As an illustrative example, we consider a theory with the gauge group SU($N_c$)
that contains $N_f \equiv N_F$ fermions in the fundamental ($F$) representation
and $N_{f'} \equiv N_{Adj}$ fermions in the adjoint representation, $Adj$. We
denote this as the FA theory.  Here the upper boundary of the NACP region 
${\cal R}_{NACP}$, which is also the upper boundary of the region ${\cal
  R}_{IRZ}$, is given by the line 
\beq
{\rm FA \ theory}: \quad N_F + 2N_c N_{Adj} = \frac{11N_c}{2} \ . 
\label{nacp_upper_fund_adj}
\eeq
Thus, $N_F < (11/2)N_c$ if $N_{Adj}=0$ and $N_{Adj} < 11/4 = 2.75$ if $N_f=0$. 
The lower boundary of ${\cal R}_{IRZ}$, which can provide an approximate
estimate to the lower boundary of ${\cal R}_{NACP}$, is given by the line 
$b_2=0$, namely 
\beqs
&& {\rm FA \ theory}: \ \ (13N_c - 3N_c^{-1})N_F + 32N_c^2 N_{Adj} = 34N_c^2
\ . \cr\cr
&& 
\label{alfir_2loop_lower_fund_adj}
\eeqs
Thus, in ${\cal R}_{IRZ}$, it follows that $N_F > 34N_c^3/(13N_c^2-3)$ if 
$N_{Adj}=0$ and $N_{Adj} > 17/16 = 1.0625$ if $N_F=0$. In this FA theory, 
the line $b_1=0$ has slope
\beq
{\rm FA \ theory}: \quad 
\frac{\partial N_{Adj}}{\partial N_F}{}\Bigg |_{b_1=0} = -\frac{1}{2N_c} \ , 
\label{b1zslope_fadj}
\eeq
while the line $b_2=0$ has slope
\beq
{\rm FA \ theory}: \quad 
 \frac{\partial N_{Adj}}{\partial N_F}{}\Bigg |_{b_2=0} = 
-\frac{(13N_c^2-3)}{32N_c^3} \ . 
\label{b2zslope_fadj}
\eeq

For example, in the FA theory with $N_c=3$, so $G={\rm SU}(3)$, these slopes
(\ref{b1zslope_fadj}) and (\ref{b2zslope_fadj}) are $-1/6=-0.16667$ and
$-19/144=-0.13194$, respectively, where the floating-point values are given to
the indicated accuracy. The $b_1=0$ line crosses the horizontal and vertical
axes at $(N_f,N_{f'})=(16.5,0)$ and $(0,2.75)$, respectively, while the $b_2=0$
line crosses the horizontal and vertical axes at $(N_f,N_{f'})=(8.0526,0)$ and
$(0,1.0625)$, respectively. 

In Table \ref{fadj_su3_theories_table} we list the physical integral values of
$N_F$ and integral and half-integral (Majorana) values of $N_{Adj}$ in the
region ${\cal R}_{IRZ}$ in this SU(3) theory. Considering $(N_F,N_{Adj})$ as a
point in the first quadrant of an ${\mathbb R}^2$ space, we list in the second
column the distance $d_u$ of this point from the line $b_1=0$ that forms the
upper boundary of the regions ${\cal R}_{IRZ}$ and ${\cal R}_{NACP}$, and in
the third column the distance $d_\ell$ of this point from the line $b_2=0$ that
forms the lower boundary of the region ${\cal R}_{IRZ}$. (By distance of a
point $P$ from a line $L$, we mean the length of the line segment perpendicular
to the line $L$ that passes through the point $P$.)  Thus, Ttable
\ref{fadj_su3_theories_table} provides a guide to the position of a theory with
a given set of values of $(N_F,N_{Adj})$ in the region ${\cal R}_{IRZ}$. In
general, theories with small values of $d_u$ are close to the upper boundary of
the region ${\cal R}_{NACP}$ and have correspondingly small
values of $\alpha_{IR}$. In order for our perturbative analysis to be
self-consistent, it is necessary that $\alpha_{IR}$ should not be
excessively large, and so one may require, say, that $\alpha_{IR,2\ell} < 1$.
Our perturbative analysis is expected to be most accurate for the
$(N_F,N_{Adj})$ FA theories with small $d_u$ and hence small
$\alpha_{IR,2\ell}$ in the upper part of the NACP. We will discuss this
illustrative two-representation FA theory further below.  


\section{Scheme-Independent Calculation of Anomalous Dimensions of Fermion
  Bilinear Operators}
\label{kappa_section}

In this section, for a theory with a general gauge group $G$ containing 
$N_f$ fermions in a representation $R$ and $N_{f'}$ fermions in a
representation $R'$, we present our new calculations of the coefficients
$\kappa^{(f)}_j$ and $\kappa^{(f')}_j$ in the scheme-independent expansions
of the anomalous dimensions $\gamma_{\bar\psi\psi,IR}$ and 
$\gamma_{\bar\chi\chi,IR}$ in Eqs. (\ref{gamma_ir_Deltaseries}) and the
analogue for $\gamma_{\bar\chi\chi,IR}$ with $1 \le j \le 3$.  
It will be useful to define a factor that occurs repeatedly in the 
denominators of various expressions, namely 
\beq
{\cal D}_f = C_A(7C_A+11C_f) + 4N_{f'}T_{f'}(C_{f'}-C_f)  \ . 
\label{dcal}
\eeq
In the previously studied theory with a single fermion representation, i.e.,
$N_{f'}=0$, this factor ${\cal D}$ reduces as
\beq
{\cal D}_f = C_AD \quad {\rm if} \ \ N_{f'}=0 \ , 
\label{dcad}
\eeq
where 
\beq
D=7C_A+11C_f \ , 
\label{dfac}
\eeq
as defined in Eq. (2.13) of our earlier work \cite{dex,dexl}.

For the first two coefficients we calculate 
\beq
\kappa^{(f)}_1 = \frac{8C_fT_f}{{\cal D}_f} 
\label{kappa1}
\eeq
and
\begin{widetext} 
\beq
\kappa^{(f)}_2 = \frac{4C_fT_f^2}{3{\cal D}_f^3} \, \Bigg [ 
C_A(7C_A+4C_f)(5C_A+88C_f) + 2^4 N_{f'}T_{f'}(C_{f'}-C_f)
\Big (10C_A+8C_f+C_{f'}\Big ) \Bigg ] \ . 
\label{kappa2}
\eeq
For the third coefficient, we write 
\beq
\kappa^{(f)}_3 = \frac{4C_fT_f}{3^4 {\cal D}_f^5} 
\Bigg [ A^{(f)}_0 + A^{(f)}_1 N_{f'} + A^{(f)}_2 N_{f'}^2 
+ A^{(f)}_3  N_{f'}^3 \Bigg ] \ . 
\label{kappa3}
\eeq
It follows that the $A^{(f)}_0$ term is independent of $N_{f'}$ and hence,
taking into account the difference in the prefactor, it is equal to $C_A$ times
the terms in the square bracket of Eq. (6.7) in our earlier
Ref. \cite{dex} or equivalently Eq. (3.4) in our Ref. \cite{dexl}. 
We have
\beqs
A^{(f)}_0 &=& C_A \Bigg [ 3C_AT_f^2 \bigg ( -18473C_A^4 + 144004 C_A^3C_f
+650896C_A^2C_f^2 +356928C_AC_f^3+569184C_f^4 \bigg ) \cr\cr
&+&2^7D\bigg (-20T_f^2\frac{d_A^{abcd}d_A^{abcd}}{d_A}
+352C_AT_f\frac{d_f^{abcd}d_A^{abcd}}{d_A}
-1331C_A^2\frac{d_f^{abcd}d_f^{abcd}}{d_A} \bigg ) \cr\cr
&+& 33 \cdot 2^{10}D \zeta_3 \, \bigg (2T_f^2 \frac{d_A^{abcd}d_A^{abcd}}{d_A}
-13 C_A T_f \frac{d_f^{abcd}d_A^{abcd}}{d_A}
+11 C_A^2 \frac{d_f^{abcd}d_f^{abcd}}{d_A} \bigg )  \ \Bigg ] \ \ ,
\label{kappa3_A0}
\eeqs
where $\zeta_s = \sum_{n=1}^\infty n^{-s}$ is the Riemann zeta function. Here,
the group invariants $C_A$, $C_f$, $T_f$, $d_A^{abcd}d_A^{abcd}$,
$d_f^{abcd}d_A^{abcd}$, $d_f^{abcd}d_f^{abcd}$, and
$d_f^{abcd}d_{f'}^{abcd}$ are defined in Appendix
\ref{groupinvariants}, and $d_A$ is the dimension of the adjoint representation
of $G$.

For the other $A_s^{(f)}$ with $1 \le s \le 3$, we calculate
\beqs
A^{(f)}_1 &=& C_AT_f^2 T_{f'}\Bigg [ 273840C_A^3(C_{f'}-C_f) 
+C_A^2 \bigg ( -1511040C_f^2 + 1916256C_f C_{f'} -405216 C_{f'}^2 \bigg ) 
\cr\cr
&&+C_A\bigg ( -129600C_f^3+522432C_f^2 C_{f'}-485568C_f C_{f'}^2 
+ 92736C_{f'}^3 \bigg ) 
\cr\cr
&&+C_f \bigg (-1241856C_f^3 +1020096C_f^2C_{f'} + 76032C_f C_{f'}^2
+145728C_{f'}^3 \bigg ) 
\ \Bigg ] \cr\cr
&&+10240 T_f^2 T_{f'}(C_f-C_{f'}) \frac{d_A^{abcd}d_A^{abdc}}{d_A}
+C_AT_f T_{f'}  \bigg ( -114688C_A -360448C_f +180224C_{f'} \bigg ) 
\frac{d_f^{abcd}d_A^{abcd}}{d_A} 
 \cr\cr
&&+C_AT_f^2 \Big ( 114688C_A + 180224 C_f \Big ) 
\frac{d_{f'}^{abcd}d_A^{abcd}}{d_A}
+C_A^2 T_{f'}\bigg ( 867328 C_A + 2044416 C_f -681472 C_{f'} \bigg )
\frac{d_f^{abcd}d_f^{abcd}}{d_A} \cr\cr
&&+C_A^2T_f\Big (-867328C_A-1362944C_f\Big ) 
\frac{d_f^{abcd}d_{f'}^{abcd}}{d_A} \cr\cr
&&+\zeta_3  \Bigg [ 270336 T_f^2 T_{f'}(C_{f'} - C_f)
  \frac{d_A^{abcd}d_A^{abcd}}{d_A}
+ C_AT_fT_{f'} \bigg ( 1118208 C_A + 3514368 C_f -1757184C_{f'} \bigg )
\frac{d_f^{abcd}d_A^{abcd}}{d_A} \cr\cr
&&+C_A T_f^2 \Big ( -1118208C_A-1757184C_f \Big )
\frac{d_{f'}^{abcd}d_A^{abcd}}{d_A} 
+C_A^2T_{f'} \bigg (-1892352 C_A -4460544 C_f +1486848C_{f'} \bigg ) 
\frac{d_f^{abcd}d_f^{abcd}}{d_A} \cr\cr
&&+C_A^2T_f \Big ( 1892352C_A + 2973696C_f \Big )
\frac{d_f^{abcd}d_{f'}^{abcd}}{d_A} 
 \Bigg ]
\label{kappa3_A1}
\eeqs
\beqs
A^{(f)}_2 &=& T_f^2 T_{f'}^2 \Bigg [ 350976 C_A^2(C_f-C_{f'})^2  
+ C_A \bigg (-94464 C_f^3 -2304C_f^2C_{f'} +288000C_f C_{f'}^2 
- 191232C_{f'}^3 \bigg ) \cr\cr
&& +225792C_f^4 -370944C_f^3C_{f'} + 119808C_f^2C_{f'}^2 
-29952C_f C_{f'}^3 + 55296C_{f'}^4 \Bigg ] \cr\cr
&&+ 2^{16}T_fT_{f'}(C_f-C_{f'})( T_{f'} \frac{d_f^{abcd}d_A^{abcd}}{d_A} 
- T_f \frac{d_{f'}^{abcd}d_A^{abcd}}{d_A}) \cr\cr
&&+C_A T_{f'}^2 \bigg ( -157696C_A+495616C_{f'}-743424C_f \bigg )
\frac{d_f^{abcd}d_f^{abcd}}{d_A}  \cr\cr
&&+C_AT_fT_{f'}\bigg ( 315392C_A+991232C_f-495616C_{f'} \bigg )
\frac{d_f^{abcd}d_{f'}^{abcd}}{d_A} 
+C_A T_f^2 ( -157696C_A -247808C_f ) 
\frac{d_{f'}^{abcd}d_{f'}^{abcd}}{d_A}
\cr\cr
&&+ \zeta_3 \Bigg [ 638976T_f T_{f'}(C_f-C_{f'})\bigg ( 
T_f \frac{d_{f'}^{abcd}d_A^{abcd}}{d_A} 
-T_{f'} \frac{d_f^{abcd}d_A^{abcd}}{d_A} \bigg ) \cr\cr
&&+C_AT_{f'}^2\bigg ( 344064C_A +1622016C_f - 1081344C_{f'} \bigg )
\frac{d_f^{abcd}d_f^{abcd}}{d_A} \cr\cr
&&C_AT_f T_{f'}\bigg ( -688128C_A -2162688C_f +1081344C_{f'} \bigg ) 
\frac{d_f^{abcd}d_{f'}^{abcd}}{d_A} 
+ C_A T_f^2 \Big ( 344064C_A + 540672C_f \Big )
\frac{d_{f'}^{abcd}d_{f'}^{abcd}}{d_A} \ \Bigg ] \cr\cr
&&
\eeqs
and
\beqs
A^{(f)}_3 &=& 2^{13} \, T_{f'}(C_f-C_{f'})(11-24\zeta_3) 
\bigg ( T_{f'}^2 \frac{d_f^{abcd}d_f^{abcd}}{d_A} 
- 2 T_f T_{f'} \frac{d_f^{abcd}d_{f'}^{abcd}}{d_A}
+ T_f^2 \frac{d_{f'}^{abcd}d_{f'}^{abcd}}{d_A} \bigg ) \ . \cr\cr
&&
\label{A3A3bar}
\eeqs
\end{widetext}

The coefficients $\kappa^{(f')}_j$ are obtained from these $\kappa^{(f)}_j$ by
interchanging $f$ and $f'$ in all expressions. For example, 
\beq
{\cal D}_{f'} = C_A(7C_A+11C_{f'}) + 4N_fT_f(C_f-C_{f'}) \ , 
\label{dcalf2}
\eeq
\beq
\kappa^{(f')}_1 = \frac{8C_{f'}T_{f'}}{{\cal D}_{f'}} \ , 
\label{kappa1f2}
\eeq
and so forth for the other expressions.

An important result that we found in our previous work \cite{gsi}-\cite{dexo}
was that for a theory with a single representation, $\kappa^{(f)}_1$ and
$\kappa^{(f)}_2$ are manifestly positive, and for all of the specific gauge
groups and fermion representations that we considered, $\kappa^{(f)}_3$ and
$\kappa^{(f)}_4$ are also positive.  This property implied several monotonicity
relations for our calculation of $\gamma_{\bar\psi\psi}$ to maximal power
$\Delta_f^p$, denoted $\gamma_{\bar\psi\psi,\Delta_f^p}$, namely that (i) for
fixed $p$, $\gamma_{\bar\psi\psi,\Delta_f^p}$ is a monotonically increasing
function of $\Delta_f$, i.e., a monotonically increasing function of decreasing
$N_f$, and (ii) for fixed $N_f$, $\gamma_{\bar\psi\psi,\Delta_f^p}$ is a
monotonically increasing function of the maximal power $p$.

A basic question that we may ask concerning these results is how a coefficient
$\kappa^{(f)}$ changes as one goes from the single-representation theory with
${N_f'}=0$ to theories with an increasing number $N_{f'}$ of fermions in a
different representation, and vice versa for the dependence of $\kappa^{(f')}$
on $N_f$.  For the purpose of this discussion, we recall that, by convention,
we take $f$ to be the fermion in the representation with a smaller dimension.
In the cases with which we deal, this also means that $C_f < C_{f'}$.  The
question is readily answered in the case of $\kappa_1^{(f)}$ and
$\kappa_1^{(f')}$. As a lemma, we observe that ${\cal D}_f$ is a monotonically
increasing function of $N_{f'}$, while ${\cal D}_{f'}$ is a monotonically
decreasing function of $N_f$. Hence, $\kappa_1^{(f)}$ is a monotonically
decreasing function of $N_{f'}$, while $\kappa_1^{(f')}$ is a monotonically
increasing function of $N_f$.  The dependence of $\kappa^{(f)}_j$ on $N_{f'}$
and of $\kappa^{(f')}_j$ on $N_f$ for indices $j=2,3$ will be analyzed below
for particular theories.

Concerning the question of the positivity of $\kappa^{(f)}_j$ and
$\kappa^{(f')}_j$, in a theory with fermions in multiple different
representations, there are terms of both signs in the expressions for the
coefficients $\kappa^{(f)}_j$.  Nevertheless, anticipating our results below,
in the specific FA theories that we have studied in detail, both
$\kappa^{(F)}_j$ and $\kappa^{(Adj)}_j$ are positive for all of the orders that
we have calculated, namely $j=1,2,3$. 

In our earlier work \cite{gtr}-\cite{pgb} on scheme-independent series
calculations for theories with a $N_f$ fermions transforming according to a
single type of representation, we carried out detailed studies of the
reliability of these expansions using a variety of methods.  One of the
simplest procedures is to analyze the fractional change in a quantity,
calculated to a given order $O(\Delta_f^p)$, as one increases the maximal power
$p$ of the expansion.  Here we shall apply this method in our illustrative
theory discussed in the next section.


\section{Anomalous Dimensions in a Theory with Fermions in the Fundamental 
and Adjoint Representations of SU($N_c$)}
\label{kappa_fadj_section}

In this section we discuss our scheme-independent calculations of 
$\gamma_{\bar\psi\psi,IR}$ and $\gamma_{\bar\chi\chi,IR}$ for the illustrative
case of a theory with gauge group SU($N_c$) containing $N_f \equiv N_F$ 
fermions in the fundamental representation and $N_{f'} \equiv N_{Adj}$ fermions
in the adjoint representation.  As before, we call this the FA theory. 
In this case, the denominator factor ${\cal D}_f$ takes the form
\beq
{\rm FA \ theory}: \quad {\cal D}_f = \frac{1}{2}\bigg [ 25N_c^2-11+
4N_{Adj}(N_c^2+1) \bigg ] \ . 
\label{dcal_fa}
\eeq
We have given the values of $(N_F,N_{Adj})$
in Table \ref{fadj_su3_theories_table} for the region ${\cal R}_{IRZ}$. 
For the first-order coefficients we calculate 
\beq
\kappa_1^{(F)} = \frac{4(N_c^2-1)}
{N_c \Big [ 25N_c^2-11 + 4N_{Adj}(N_c^2+1) \Big ]}
\label{kappa1_fund_fa}
\eeq
and
\beq
\kappa_1^{(Adj)} = \frac{8N_c^3}{18N_c^3 - N_F(N_c^2+1)} \ . 
\label{kappa1_adj_fa}
\eeq
If $N_{Adj}=0$, then the coefficient $\kappa_1^{(F)}$ reduces to the expression
$4(N_c^2-1)/[N_c(25N_c^2-11)]$, as given in Eq. (6.8) of our earlier work
\cite{dex}.  Similarly, if $N_F=0$, then $\kappa_1^{(Adj)}$ reduces to the
value 4/9, as given in Eq. (6.18)) of \cite{dex}.

For the second-order coefficients, we find 
\begin{widetext}
\beq
\kappa_2^{(F)} =  \frac{4(N_c^2-1)\Big [ (9N_c^2-2)(49N_c^2-44) + 
8N_{Adj}(N_c^2+1)(15N_c^2-4) \, \Big ]}
{3N_c^2\Big [ 25N_c^2-11 + 4N_{Adj}(N_c^2+1) \Big ]^3}
\label{kappa2_fund_fa}
\eeq
and
\beq
\kappa_2^{(Adj)} = \frac{4N_c^4\Big [ 1023N_c^5
-2N_F(N_c^2+1)(37N_c^2-1)  \, \Big ] }
{3\Big [ 18N_c^3-N_F(N_c^2+1) \Big ]^3} \ . 
\label{kappa2_adj_fa}
\eeq
\end{widetext}
If $N_{Adj}=0$, then $\kappa_2^{(F)}$ reduces to the expression
given in Eq. (6.9) of \cite{dex}, and if $N_F=0$, then $\kappa_2^{(Adj)}$
reduces to the value $341/1458 = 341/(2 \cdot 3^6)$ as given in Eq. (6.19) of
\cite{dex}.

Our results for the third-order coefficients are as follows:
\begin{widetext}
\beq
\kappa_3^{(F)} =  \frac{2^3(N_c^2-1)\Big [ \kappa^{(F)}_{3,0} + 
\kappa^{(F)}_{3,1}N_{Adj}  + \kappa^{(F)}_{3,2}N_{Adj}^2  + 
\kappa^{(F)}_{3,3}N_{Adj}^3  \ \Big ]}
{3^3 N_c^3\Big [ 25N_c^2-11 + 4N_{Adj}(N_c^2+1) \, \Big ]^5 } \ , 
\label{kappa3_fund_fa}
\eeq
where 
\beqs
\kappa^{(F)}_{3,0} &=& 
 274243N_c^8-455426N_c^6-114080N_c^4+47344N_c^2+35574 
\cr\cr
&-& 2^7 \cdot 33 \zeta_3 N_c^2(4N_c^2-11)(25N_c^2-11)
\label{kappabar_fund_fa_Nadjoint0}
\eeqs
\beqs
\kappa^{(F)}_{3,1} &=&  135848N_c^8 - 215832N_c^6 + 291424N_c^4 
-189168N_c^2 -25872  \cr\cr
&-& 2^9 \cdot 3^2 \zeta_3 N_c^2 \Big ( 73N_c^4+132N_c^2-121 \Big ) 
\label{kappabar_fund_fa_Nadjlinear}
\eeqs
\beqs
\kappa^{(F)}_{3,2} &=& 2^5(N_c^2+1)\Big [ 689N_c^6-2651N_c^4+2775N_c^2
+147 + 2^6 \cdot 3^2 \zeta_3 N_c^2(6N_c^2 -11) \, \Big ] 
\label{kappabar_fund_fa_Nadjoint_quadratic}
\eeqs
and
\beq
\kappa^{(F)}_{3,3} = 2^{10}N_c^2(N_c^2+1)^2(-11+24\zeta_3) \ . 
\label{kappabar_fund_fa_Nadjcubic}
\eeq
Further, 
\beq
\kappa_3^{(Adj)} =  \frac{4N_c^5\Big [ \kappa^{(Adj)}_{3,0} + 
\kappa^{(Adj)}_{3,1}N_F  + \kappa^{(Adj)}_{3,2}N_F^2  + 
\kappa^{(Adj)}_{3,3}N_F^3 \ \Big ]}
{3^3\Big [18N_c^3-N_F(N_c^2+1) \Big ]^5} \ , 
\label{kappa3_adj_fa}
\eeq
where 
\beq
\kappa^{(Adj)}_{3,0} = 3^3 N_c^8(61873N_c^2-42624)
\label{kappabar_adj_fa_Nfund0}
\eeq
\beq
\kappa^{(Adj)}_{3,1} = -36N_c^3\Big ( 6728N_c^6 -5857N_c^4-1247N_c^2+138
+ 11520 \zeta_3N_c^4 \Big ) 
\label{kappabar_adj_fa_Nfundlinear}
\eeq
\beqs
\kappa^{(Adj)}_{3,2} &=& 2^5(N_c^2+1)\Big ( 287N_c^6-1187N_c^4 +
27N_c^2+9 + 2448\zeta_3 N_c^4 \Big )
\label{kappabar_adj_fa_Nfundquadratic}
\eeqs
and
\beq
\kappa^{(Adj)}_{3,3} = -2^7N_c(N_c^2+1)^2(-11+24\zeta_3) \ . 
\label{kappabar_adj_fa_Nfundcubic}
\eeq
\end{widetext}
If $N_{Adj}=0$, then the coefficient $\kappa_3^{(F)}$ reduces to the expression
in Eq. (6.10) of our earlier work \cite{dex}, while if $N_F=0$, then 
$\kappa_3^{(Adj)}$ reduces to Eq. (6.20) of \cite{dex}. The agreement of these
reductions of $\kappa_j^{(F)}$ for $N_{Adj}=0$ and of $\kappa_j^{(Adj)}$ for
$N_F=0$ with our earlier calculations in \cite{dex} for $j=1,2,3$ serves as a
check on our present results.  As was discussed in \cite{dex,dexl}, these
coefficients have the leading large-$N_c$ dependence 
\beq
\kappa_j^{(F)} \sim N_c^{-j} \quad {\rm as} \ N_c \to \infty
\label{kappa_fund_largeNc}
\eeq
and
\beq
\kappa_j^{(Adj)} \sim N_c^0 \quad {\rm as} \ N_c \to \infty \ . 
\label{kappa_adj_largeNc}
\eeq

As specific examples of these FA theories, we consider the following sets of
SU(3) gauge theories in ${\cal R}_{IRZ}$ with the indicated fermion content:
\beqs
(N_F,N_{Adj}) &=&  (8,\frac{1}{2}), \quad (8,1), \quad (10,0), 
\quad (10,\frac{1}{2}), \cr\cr
&& (10,1), \quad (12,0), \quad (12,\frac{1}{2}) \ . 
\label{fa_theories}
\eeqs
The respective positions of these theories in the regions ${\cal R}_{IRZ}$ and
${\cal R}_{NACP}$ can be ascertained by referring to Table
\ref{fadj_su3_theories_table}.  The corresponding values of the coefficients
$\kappa^{(F)}_j$ with $j=1,2,3$, as functions of $N_{Adj}$, are listed in Table
\ref{kappa_fund_su3fa_table}, and the values of $\kappa^{(Adj)}_j$ with
$j=1,2,3$, as functions of $N_F$, are listed in Table
\ref{kappa_adj_su3fa_table}.

We observe that all of these coefficients are positive, and so the
generalizations of the monotonicity relations that we found in our earlier work
for the theory with fermions in a single representation also hold for this FA
theory, namely (i) for fixed $N_{Adj}$, $\gamma_{\bar\psi\psi,IR}$ is a
monotonically increasing function of $\Delta_F$, i.e., a monotonically
increasing function of decreasing $N_F$; (ii) for fixed $N_F$,
$\gamma_{\bar\chi\chi,IR}$ is a monotonically increasing function of
$\Delta_{Adj}$, i.e., a monotonically increasing function of decreasing
$N_{Adj}$; (iii) for fixed $N_{f'}$, $\gamma_{\bar\psi\psi,IR,\Delta_F^p}$ is a
monotonically increasing function of $p$; and (iv) for fixed $N_f$,
$\gamma_{\bar\chi\chi,IR,\Delta_{Adj}^p}$ is a monotonically increasing
function of $p$.

Separately, we also note a generalization of the monotonicity relation that we
proved for $\kappa^{(F)}_1$ and proved for $\kappa^{(Adj)}_1$, namely that for
these FA theories, the $\kappa^{(F)}_j$ coefficients with $j=1,2,3$ 
are monotonically decreasing functions of
$N_{Adj}$, and the $\kappa^{(Adj)}_j$ coefficients with $j=1,2,3$ 
are monotonically increasing functions of $N_F$.

Having calculated these coefficients $\kappa^{(F)}_j$ and 
$\kappa^{(Adj)}_j$ with $j=1,2,3$ for this FA theory, we next proceed to
substitute them in the general scheme-independent expansions 
(\ref{gamma_ir_Deltaseries}) for $f=F$ and the analogue for $f'=Adj$. 
Explicitly, with $f = \psi$ and $f'=\chi$, 
\beq
\gamma_{\bar\psi\psi,IR} = \sum_{j=1}^\infty \kappa^{(F)}_j \Delta_F^j
\label{gamma_fund_Deltaseries_fa}
\eeq
and
\beq
\gamma_{\bar\chi\chi,IR} = \sum_{j=1}^\infty \kappa^{(Adj)}_j \Delta_{Adj}^j \ ,
\label{gamma_adj_Deltaseries_fa}
\eeq
where
\beq
\Delta_F = (N_{F,u}-N_F) \ , 
\label{Delta_fund}
\eeq
with 
\beq
N_{F,u} = \frac{N_c(11-4N_{Adj})}{2}
\label{Nfundup}
\eeq
and
\beq
\Delta_{Adj} = (N_{Adj,u}-N_{Adj}) \ , 
\label{Delta_Adj}
\eeq
with
\beq
N_{Adj,u} = \frac{11}{4} - \frac{N_F}{2N_c} \ . 
\label{Nadjup}
\eeq
For reference, we list the values of $N_{F,u}$ and $N_{Adj,u}$ from 
Eqs. (\ref{Nfundup}) and (\ref{Nadjup}) for these $(N_F,N_{Adj})$ FA SU(3) 
theories in Table \ref{nup_values_su3_fa}. 

In Table \ref{gamma_fund_fa_su3_values} we list the values of
$\gamma_{\bar\psi\psi,IR}$ calculated to $O(\Delta_F^p)$ for $p=1,2,3$, denoted
as $\gamma_{\bar\psi\psi,IR,\Delta_F^p}$.  Similarly, 
in Table \ref{gamma_adj_fa_su3_values} we list the values of
$\gamma_{\bar\chi\chi,IR}$ calculated to $O(\Delta_{Adj}^p)$ for $p=1,2,3$, 
denoted as $\gamma_{\bar\chi\chi,IR,\Delta_{Adj}^p}$. The monotonicity
relations noted above are evident in these tables.  From an examination of the
fractional changes in the anomalous dimensions as one increases the order of
calculation, one may infer that these scheme-independent expansions should be
reasonably reliable.  For example, in the SU(3) FA theory with
$(N_F,N_{Adj})=(12,1/2)$ theory, the fractional change in the
$\gamma_{\bar\psi\psi,IR}$ anomalous dimension is 
\begin{widetext}
\beq
{\rm SU}(3), \ (N_F,N_{Adj}) = (12,1/2) \Longrightarrow \quad 
\frac{\gamma_{\bar\psi\psi,IR,\Delta_F^3} - 
\gamma_{\bar\psi\psi,IR,\Delta_F^2}}{\gamma_{\bar\psi\psi,IR,\Delta_F^2}} =
0.81 \times 10^{-2} \ . 
\eeq
In the SU(3) FA theory with $(N_F,N_{Adj})=(10,1)$ theory, the fractional
change in $\gamma_{\bar\psi\psi,IR}$ is even smaller:
\beq
{\rm SU}(3), \ (N_F,N_{Adj}) = (10,1) \Longrightarrow \quad 
\frac{\gamma_{\bar\psi\psi,IR,\Delta_F^3} - 
\gamma_{\bar\psi\psi,IR,\Delta_F^2}}{\gamma_{\bar\psi\psi,IR,\Delta_F^2}} =
0.87 \times 10^{-3} \ , 
\eeq
yielding identical entries listed to three significant figures in Table
\ref{gamma_fund_fa_su3_values}.  Similar comments apply to the
calculations of $\gamma_{\bar\chi\chi,IR,\Delta_{Adj}^p}$. 
\end{widetext}


\section{Scheme-Independent Calculation of $\beta'_{IR}$}
\label{betaprime_section}

In this section we return to the general asymptotically free gauge theory with
gauge group $G$ containing $N_f$ and $N_{f'}$ fermions in the respective
representations $R$ and $R'$ and present our calculations of the coefficients
$d_j$ and $\tilde d_j$ in the scheme-independent expansions of the derivative 
of the beta function evaluated at the IR fixed point, $\beta'_{IR}$, 
in powers of $\Delta_f$ in Eqs. (\ref{betaprime_ir_Deltaseries}) and in powers
of $\Delta_{f'}$ in Eq. (\ref{betaprime_ir_Deltaseries2}), respectively.  
As before in this paper,
this IR fixed point is taken to be in the non-Abelian Coulomb phase. Part of
the physical interest in the quantity $\beta'_{IR}$ stems from the fact that,
owing to the trace anomaly relation \cite{traceanomaly}, it is equivalent to
the anomalous dimension of the field-strength tensor term ${\rm
  Tr}(F^a_{\mu\nu}F^{a \mu\nu})$ in the Lagrangian \cite{tracerel,dex}.
As noted above, generalizing our result for the single-representation case,
$d_1 = \tilde d_1=0$ for arbitrary $G$, $R$, and $R'$.

For the higher coefficients we find
\beq
d_2 = \frac{2^5 T_f^2}{3^2 {\cal D}_f}
\label{d2}
\eeq
\beq
d_3 = \frac{2^7 T_f^3 (5C_A+3C_f)}{3^3 {\cal D}_f^2} 
\label{d3}
\eeq
and
\beq
d_4 = -\frac{2^3T_f^2}{3^6{\cal  D}_f^5} \, 
\Big [ B^{(f)}_0 + B^{(f)}_1 N_{f'} + B^{(f)}_2 N_{f'}^2  
+  B^{(f)}_3 N_{f'}^3 \Big ] \ , 
\label{d4}
\eeq
where we explicitly indicate the dependence on $f$ in the $B^{(f)}_s$,
$s=0,1,2,3$.  
(We extract a minus sign in Eq. (\ref{d4}) to maintain the same notation as in
our earlier works \cite{dex,dexl}, where we found that in the case of fermions
in a single representation $R=F$, $d_4$ is negative.) 
As was the case with $A^{(f)}_0$ in $\kappa^{(f)}_3$, the $B^{(f)}_0$ term in 
$d_4$ is independent of $N_{f'}$ and hence, taking into
account the difference in the prefactor, it is equal to $C_A$ times 
the terms in the square bracket of Eq. (5.11) in our earlier
Ref. \cite{dex} or equivalently, Eq. (4.8) of our Ref. \cite{dexl}. We have
\begin{widetext} 
\beqs
B^{(f)}_0 &=& C_A \Bigg [ 
-3C_AT_f^2 \bigg ( 137445 C_A^4 + 103600C_A^3C_f+72616C_A^2C_f^2
+951808C_AC_f^3 - 63888C_f^4 \bigg ) \cr\cr
&+&2^8D \bigg ( -20T_f^2 \frac{d_A^{abcd}d_A^{abcd}}{d_A}
+352C_AT_f\frac{d_f^{abcd}d_A^{abcd}}{d_A}
-1331C_A^2\frac{d_f^{abcd}d_f^{abcd}}{d_A} \bigg ) \cr\cr
&+& 8448D \zeta_3 \, \bigg \{ C_A^2T_f^2\Big ( 21C_A^2+12C_AC_f-33C_f^2\Big )
+ 16T_f^2 \frac{d_A^{abcd}d_A^{abcd}}{d_A}
- 104 C_AT_f \frac{d_f^{abcd}d_A^{abcd}}{d_A}
+ 88C_A^2\frac{d_f^{abcd}d_f^{abcd}}{d_A} \bigg \} \ \Bigg ] \ . \cr\cr
& &
\label{d4_B0}
\eeqs
For the $B^{(f)}_j$ with $j=1,2,3$, we calculate 
\beqs
B^{(f)}_1 &=& 194880C_A^4T_f^2T_{f'}(C_{f'}-C_f) +
C_A^3T_f^2T_{f'}\bigg ( -1854816C_f^2+2715648C_fC_{f'}-860832C_{f'}^2\bigg )
\cr\cr
&+& C_A^2T_f^2T_{f'}\bigg ( 903168C_f^3+153216C_f^2C_{f'}
-1241856C_fC_{f'}^2+185472C_{f'}^3 \bigg ) \cr\cr
&+&C_AT_f^2T_{f'}C_f \bigg ( -139392C_f^3-164736C_f^2C_{f'}
+12672C_fC_{f'}^2+291456C_{f'}^3 \bigg ) \cr\cr
&+& C_A^2T_f^2T_{f'} (C_f-C_{f'}) \zeta_3 \, \bigg ( -967680C_A^2 +608256C_AC_f +
3345408C_f^2 \bigg ) \cr\cr
&+& T_f^2T_{f'}(C_f-C_{f'})\Big (20480-540672\zeta_3\Big )
\frac{d_A^{abcd}d_A^{abcd}}{d_A} \cr\cr
&+& C_AT_fT_{f'}\bigg [ -229376C_A-720896C_f+360448C_{f'} + 
\zeta_3\Big ( 2236416C_A+7028736C_f-3514368C_{f'} \Big ) \, \bigg ]\, 
\frac{d_f^{abcd}d_A^{abcd}}{d_A} \cr\cr
&+&C_AT_f^2 \bigg [ 229376C_A+360448C_f + \zeta_3\bigg
  (-2236416C_A-3514368C_f\bigg ) \,  \bigg ]\frac{d_{f'}^{abcd}d_A^{abcd}}{d_A} \cr\cr
&+&C_A^2T_{f'}\bigg [ 1734656C_A+4088832C_f-1362944C_{f'} + 
\zeta_3\bigg ( -3784704C_A-8921088C_f+2973696C_{f'}\bigg ) \, \bigg ]
\frac{d_f^{abcd}d_f^{abcd}}{d_A} \cr\cr
&+&C_A^2T_f \bigg [ C_A\Big ( -1734656+3784704\zeta_3\Big ) + 
C_f\Big ( -2725888+5947392\zeta_3\Big ) \, \bigg ]
\frac{d_f^{abcd}d_{f'}^{abcd}}{d_A} 
\label{d4_B1}
\eeqs
\beqs
B^{(f)}_2 &=& T_f^2T_{f'}^2 \Bigg [ 669696 C_A^2(C_f-C_{f'})^2 
+C_A\bigg (437760C_f^3-1327104C_f^2C_{f'}+1340928C_fC_{f'}^2-451584C_{f'}^3\bigg) \cr\cr
&+&
25344C_f^4+59904C_f^3C_{f'}-87552C_f^2C_{f'}^2-105984C_fC_{f'}^3+108288C_{f'}^4
\cr\cr
&+& C_A(C_f-C_{f'})^2\, \zeta_3 \, \Big (-110592C_A-1216512C_f \Big ) \, \Bigg
] \cr\cr
&+& T_fT_{f'}(C_f-C_{f'})\Big (131072-1277952\zeta_3\Big )\bigg [
  T_{f'}\frac{d_f^{abcd}d_A^{abcd}}{d_A}
  -T_f\frac{d_{f'}^{abcd}d_A^{abcd}}{d_A} \bigg ] \cr\cr
&+&C_AT_{f'}^2\bigg [ -315392C_A-1486848C_f+991232C_{f'} + 
\zeta_3\bigg (688128C_A+3244032C_f-2162688C_{f'}\bigg ) \, \bigg
]\frac{d_f^{abcd}d_f^{abcd}}{d_A} \cr\cr
&+&C_AT_fT_{f'}\bigg [ 630784C_A+1982464C_f-991232C_{f'} + \zeta_3\bigg (
  -1376256C_A - 4325376C_f+2162688C_{f'}\bigg ) \, \bigg ]\frac{d_f^{abcd}d_{f'}^{abcd}}{d_A} \cr\cr
&+&C_AT_f^2\bigg [-315392C_A-495616C_f + \zeta_3\Big (688128C_A+1081344C_f\Big
  ) \, \bigg ]\frac{d_{f'}^{abcd}d_{f'}^{abcd}}{d_A} 
\label{d4_B2}
\eeqs
and
\beqs
B^{(f)}_3 &=& T_{f'}(C_f-C_{f'})\Bigg [ 2^{11} \cdot 3
  T_f^2T_{f'}^2(C_f-C_{f'})^2\Big (-23 + 24\zeta_3\Big ) \cr\cr
&&+2^{14}(11-24\zeta_3)\bigg (T_{f'}^2 \frac{d_f^{abcd}d_f^{abcd}}{d_A}
-2T_fT_{f'}\frac{d_f^{abcd}d_{f'}^{abcd}}{d_A} +
T_f^2\frac{d_{f'}^{abcd}d_{f'}^{abcd}}{d_A} \bigg ) \, \Bigg ] \ .
\label{d4_B3}
\eeqs
In passing, we note that $B^{(f)}_3$ has the same prefactor as
$A^{(f)}_3$ in Eq. (\ref{A3A3bar}), namely $T_{f'}(C_f-C_{f'})$.
\end{widetext}

The corresponding coefficients for the expansion
(\ref{betaprime_ir_Deltaseries2}) are obtained from these by
interchanging $f$ and $f'$.  Thus, for example, 
\beq
\tilde d_2 = \frac{2^5 T_{f'}^2}{3^2 {\cal D}_{f'}}
\label{d2f2}
\eeq
\beq
\tilde d_3 = \frac{2^7 T_{f'}^3 (5C_A+3C_{f'})}{3^3 {\cal D}_{f'}^2} \ , 
\label{d3f3}
\eeq
and similarly for $\tilde d_4$. 

%

\section{Results for $\beta'_{IR}$ in a Theory with Fermions in the Fundamental 
and Adjoint Representations of SU($N_c$)}
\label{betaprime_fadj_section}

In this section we discuss the special case of our general calculation of 
$\beta'_{IR}$ for an SU($N_c$) theory with $N_f$ fermions in the fundamental
representation and $N_{Adj}$ fermions in the adjoint representation (i.e., the
FA theory). We write 
Eqs. (\ref{betaprime_ir_Deltaseries}) and 
     (\ref{betaprime_ir_Deltaseries2}) as
\beq
\beta'_{IR} = \sum_{j=2}^\infty d_j \Delta_F^j 
\label{betaprime_ir_Deltaseries_fa}
\eeq
and
\beq
\beta'_{IR} = \sum_{j=2}^\infty \tilde d_j \Delta_{Adj}^j  \ , 
\label{betaprime_ir_Deltaseries2_fa}
\eeq
where $\Delta_F$ and $\Delta_{Adj}$ were given explicitly in
Eqs. (\ref{Delta_fund})-(\ref{Nadjup}).

We calculate
\beq
d_2 = \frac{2^4}{3^2\Big [ 25N_c^2-11 + 4N_{Adj}(N_c^2+1) \, \Big ]}
\label{d2_fa}
\eeq
\beq
\tilde d_2 = \frac{2^5N_c^3}{3^2\Big [18N_c^3-N_F(N_c^2+1) \, \Big ]}
\label{d2tilde_fa}
\eeq
\beq
d_3 = \frac{2^5(13N_c^2-3)}{3^3N_c\Big [ 25N_c^2-11 + 4N_{Adj}(N_c^2+1) \, 
\Big ]^2 }
\label{d3_fa}
\eeq
\beq
\tilde d_3 = \frac{2^{10}N_c^6}{3^3 \Big [18N_c^3-N_F(N_c^2+1) \, \Big ]^2} 
\label{d3tilde_fa}
\eeq
\beq
d_4 = \frac{2^4\Big [ d_{4,0} + d_{4,1} N_{Adj} + 
d_{4,2} N_{Adj}^2 + d_{4,3} N_{Adj}^3 \ \Big ]}
{3^5 N_c^2 \Big [ 25N_c^2-11 + 4N_{Adj}(N_c^2+1) \, \Big ]^5 } \ . 
\label{d4_fa}
\eeq
where 
\begin{widetext}
\beqs
d_{4,0} &=& 366782 N_c^8 -865400N_c^6+1599316N_c^4-571516N_c^2-3993
\cr\cr
&+&\zeta_3 N_c^2\Big [ -660000N_c^6 + 765600N_c^4-2241888N_c^2+894432 \Big ]
\label{d40}
\eeqs
\beqs
d_{4,1} &=& 18416N_c^8 + 346944N_c^6 - 756920N_c^4+530256N_c^2+2904 \cr\cr
&+& \zeta_3 N_c^2 \Big [ 28800N_c^6+372096N_c^4+1026432N_c^2-975744 \Big ]
\label{d4_Nadj_linear}
\eeqs
\beqs
d_{4,2} &=& 2^4(N_c^2+1)\Big [ -3161N_c^6+10589N_c^4-10155N_c^2-33 \cr\cr
&+&\zeta_3 N_c^2 ( 3744N_c^4-13248N_c^2 + 22176 ) \, \Big ] 
\label{d4_Nadj_quadratic}
\eeqs
\beqs
d_{4,3} &=& 2^8 N_c^2(N_c^2+1)^2 \Big [ -23N_c^2+65+\zeta_3(24N_c^2-168) \ , 
\, \Big ] 
\label{d4_Nadj_cubic}
\eeqs
and
\beq
\tilde d_4 = \frac{2^3 N_c^5 \Big [ \hat d_{4,0} + \hat d_{4,1} N_F + 
\hat d_{4,2} N_F^2 + \hat d_{4,3} N_F^3 \ \Big ]}
{3^5\Big [ 18N_c^3 - N_F(N_c^2+1) \, \Big ]^5 } \ , 
\label{dtilde4_fa}
\eeq
where 
\beqs
\hat d_{4,0} &=& 3^3N_c^8(46871N_c^2 + 85248) 
\label{d4hat0}
\eeqs
\beq
\hat d_{4,1} = 36 N_c^3 \Big [ 1287N_c^6-23350N_c^4-1961N_c^2+276 
+ \zeta_3 N_c^4 \Big (-6912N_c^2+16128 \Big ) \, \Big ]
\label{d4hat_NF_linear}
\eeq
\beq
\hat d_{4,2} = 4(N_c^2+1)\Big [ -5153N_c^6+18113N_c^4-747N_c^2-141
+\zeta_3 N_c^4 \Big (6912N_c^2-32256\Big ) \, \Big ] 
\label{d4hat_NF_quadratic}
\eeq
\beq
\hat d_{4,3} = 2^5 N_c(N_c^2+1)^2 \Big [ 23N_c^2 -65+\zeta_3(-24N_c^2+168)
\, \Big ] \ . 
\label{d4hat_NF_cubic}
\eeq
\end{widetext}
If $N_{Adj}=0$, then $d_2$, $d_3$, and $d_4$ reduce to our previous results in,
respectively, Eqs. (5.14), (5.15), and (5.16) of \cite{dex}.  Similarly, 
if $N_F=0$, then $\tilde d_2$, $\tilde d_3$, and $\tilde d_4$ reduce to our
previous results in, respectively, Eqs. (5.59), (5.60), and (5.61) of
\cite{dex}. The agreement of these
reductions of $d_j$ for $N_{Adj}=0$ and of $\tilde d_j$ for
$N_F=0$ with our results in \cite{dex} for $j=1,2,3$ serves as a check on our
present calculations. As was discussed in \cite{dex,dexl}, these coefficients
have the leading large-$N_c$ dependence 
\beq
d_j \sim N_c^{-j} \quad {\rm as} \ N_c \to \infty
\label{dj_largeNc}
\eeq
and
\beq
\tilde d_j \sim N_c^0 \quad {\rm as} \ N_c \to \infty \ . 
\label{dtildej_largeNc}
\eeq

In Table \ref{betaprime_fa_su3_values} we present our scheme-independent
calculations of $\beta'_{IR}$ to order $O(\Delta_F^p)$ via the expansion
(\ref{betaprime_ir_Deltaseries_fa}) and to $O(\Delta_{Adj}^p)$ via the
expansion (\ref{betaprime_ir_Deltaseries2_fa}), with $p=1,2,3$, where
$\Delta_F$ and $\Delta_{Adj}$ were defined in
Eqs. (\ref{Delta_fund})-(\ref{Nadjup}). These are denoted
$\beta'_{IR,\Delta_F^p}$ and $\beta'_{IR,\Delta_{Adj}^p}$, respectively. 
Graphically, in the first quadrant of ${\mathbb R}^2$ defined by
$(N_F,N_{adj})$ (formally generalized to non-negative real numbers), the series
(\ref{betaprime_ir_Deltaseries_fa}) is an expansion in a leftward 
horizontal direction from the $b_1=0$ line toward a given point 
$(N_F,N_{Adj})$ in the NACP, while the series 
(\ref{betaprime_ir_Deltaseries_fa}) is an expansion inward in a downward 
vertical direction from the $b_1=0$ line toward this point $(N_F,N_{Adj})$. 
Since these are two alternate expansions for the same quantity, one expects
that as the maximal power $p$ in the series increases, they should yield
similar values, and we see that this expectation is satisfied by our results at
the highest order, $p=3$, as listed in Table \ref{betaprime_fa_su3_values}. 
The agreement between the two series is best when the $(N_F,N_{Adj})$ theory is
near to the upper end of the non-Abelian Coulomb phase, since in this case the
expansion parameters $\Delta_F$ and $\Delta_{Adj}$ are the smallest.  Some 
explicit examples that demonstrate this accuracy are provided by the following
fractional differences: 
\begin{widetext}
\beq
{\rm SU}(3), \ (N_F,N_{Adj})=(10,1) \ \Longrightarrow \quad 
\frac{|\beta'_{IR,\Delta_F^4}-\beta'_{IR,\Delta_{Adj}^4}|}
{\beta'_{IR,\Delta_F^4}} = 2.2 \times 10^{-5} 
\eeq
and 
\beq
{\rm SU}(3), \ (N_F,N_{Adj})=(12,\frac{1}{2}) \ \Longrightarrow \quad 
\frac{|\beta'_{IR,\Delta_F^4}-\beta'_{IR,\Delta_{Adj}^4}|}
{\beta'_{IR,\Delta_F^4}} = 1.0 \times 10^{-3} \ . 
\eeq
\end{widetext}
%


\section{Conclusions}
\label{conclusion_section} 

In this paper, generalizing our previous work, we have considered an
asymptotically free gauge theory with gauge group $G$ and two different fermion
representations, with the property that it exhibits an infrared fixed point
such that the infrared theory is in a non-Abelian Coulomb phase. Specifically,
we have considered a theory with $N_f$ fermions transforming according to a
representation $R$ of $G$ and $N_{f'}$ fermions transforming according to a
different representation, $R'$. We have calculated scheme-independent series
expansions of the anomalous dimensions of gauge invariant fermion bilinears and
the derivative $\beta'_{IR}$ evaluated at the IR fixed point in the respective
expansion parameters $\Delta_f$ and $\Delta_{f'}$. As an explicit application,
we have presented calculations for an SU($N_c$) theory with $N_F$ fermions in
the fundamental representation and $N_{Adj}$ fermions in the adjoint
representation. Our results for scheme-independent expansions of
gauge-invariant fermion bilinears extend up to $O(\Delta_F^3)$ and
$O(\Delta_{Adj}^3)$, while our results for $\beta'_{IR}$ extend up to
$O(\Delta_F^4)$ and $O(\Delta_{Adj}^4)$. These results provide further
information about the properties of these conformal field theories.  To the
extent that the transition from the lower part of the non-Abelian Coulomb phase
to the quasi-conformal regime in the variables $(N_f,N_{f'})$ is continuous and
our finite-order perturbative calculations in the lower part of the non-Abelian
Coulomb phase are sufficiently accurate, our present results can also be useful
for the investigation of quasi-conformal theories with possible relevance to
ultraviolet completions of the Standard Model.


\begin{acknowledgments}

This research was supported in part by the Danish National
Research Foundation grant DNRF90 to CP$^3$-Origins at SDU (T.A.R.) and 
by the U.S. NSF Grant NSF-PHY-16-1620628 (R.S.) 

\end{acknowledgments}


\begin{appendix}

\section{Group Invariants}
\label{groupinvariants} 

In this appendix we discuss some relevant group-theoretic quantities. Let us 
denote the generators of the Lie algebra of the gauge group $G$, in the 
representation $R$, as $T^a_R$, with $1 \le a \le d_A$, where $d_A$ is the
order of the group. These generators satisfy the commutation relations 
\beq
[T^a_R,T^b_R]=if^{abc} T^c_R \ ,
\label{algebra}
\eeq
where the $f^{abc}$ are the associated structure constants of this Lie algebra.
Here and elsewhere, a sum over repeated indices is understood. We denote the
dimension of a given representation $R$ as $d_R = {\rm dim}(R)$. In particular,
we denote the adjoint representation by $A$, with the dimension
$d_A$ equal to the number of generators of the group, i.e., the order of the
group. The trace invariant is given by 
\beq
{\rm Tr}_R(T^a_R T^b_R) = T(R)\delta_{ab} \ .
\label{trace}
\eeq
The quadratic Casimir invariant $C_2(R)$ is defined by 
\beq
T^a_RT^a_R = C_2(R) I \ ,
\label{c2}
\eeq
where $I$ is the $d_R \times d_R$ identity matrix. For a fermion $f$
transforming according to a representation $R$, we often use the equivalent
compact notation $T_f \equiv T(R)$ and $C_f \equiv C_2(R)$. We also use the
notation $C_A \equiv C_2(A) \equiv C_2(G)$.   The invariants $T(R)$ and
$C_2(R)$ satisfy the relation $C_2(R)d_R = T(R)d_A$. For $G={\rm
SU}(N_c)$, $C_A = N_c$ and for $R$ equal to the fundamental 
representation, $T(R)=1/2$ and $C_2(R)=(N_c^2-1)/(2N_c)$. 

At the four-loop and five-loop level, one encounters traces of quartic products
of the Lie algebra generators. For a given representation $R$ of $G$,
\beqs
d^{abcd}_R &=&
\frac{1}{3!} {\rm Tr}_R\Big [ T_a(T_bT_cT_d + T_bT_dT_c + T_cT_bT_d \cr\cr
&+& T_cT_dT_b + T_dT_bT_c + T_dT_cT_b) \Big ] \ .
\label{d_abcd}
\eeqs
As with the quadratic invariants, for a fermion $f$ in the represenation $R$ of
$G$, we often use the notation $d_R^{abcd} \equiv d_f^{abcd}$. In this context,
for $R=Adj$, we use $d_R^{abcd}=d_A^{abcd}$. 
The quantities that appear in the anomalous dimensions and derivative of the
beta function $\beta'_{IR}$ that we calculate are products of these
$d_R^{abcd}$ of the form $d_R^{abcd}d_{R'}^{abcd} \equiv 
d_f^{abcd}d_{f'}^{abcd}$, summed over the group indices $a, \ b, \ c, \ d$.
For further discussion of these, with references to the literature, see 
\cite{rsv,dexo} and references therein. 

\end{appendix}



\newpage

\begin{table}
  \caption{\footnotesize{List of asymptotically free SU(3) 
gauge theories with $N_F$ fermions in the fundamental ($F$) representation and
$N_{Adj}$ fermions in the adjoint ($Adj$) representation, with the property that
the two-loop beta function has an IR zero, at $\alpha = \alpha_{IR,2\ell}$. 
The four columns list $(N_F,N_{Adj})$, $d_u$,
$d_\ell$, and $\alpha_{IR,2\ell}$, where $d_u$ and $d_\ell$ are the distances
of the point $(N_F,N_{Adj})$ to the line $b_1=0$ and to the line $b_2=0$,
respectively. Half-integral values of $N_{Adj}$ correspond to $2N_{Adj}$
copies of Majorana fermions in the adjoint representation.}}
\begin{center}
\begin{tabular}{|c|c|c|c|} \hline\hline
$(N_F,N_{Adj})$ & $d_u$ & $d_\ell$ & $\alpha_{IR,2\ell}$ 
\\ \hline
(0,3/2)  & 1.233  & 0.434  & 1.496  \\
(0,2)    & 0.740  & 0.929  & 0.419  \\
(0,5/2)  & 0.247  & 1.425  & 0.0911 \\
\hline
(1,1)    & 1.562  & 0.0688 & 11.938 \\
(1,3/2)  & 1.069  & 0.565  & 0.996  \\
(1,2)    & 0.575  & 1.060  & 0.286  \\
(1,5/2)  & 0.0822 & 1.556  & 0.0278 \\
\hline
(2,1)    & 1.397  & 0.200  & 3.683  \\
(2,3/2)  & 0.904  & 0.695  & 0.684  \\
(2,2)    & 0.411  & 1.191  & 0.182  \\
\hline
(3,1)    & 1.233  & 0.330  & 1.963  \\
(3,3/2)  & 0.740  & 0.826  & 0.471  \\
(3,2)    & 0.247  & 1.322  & 0.0982 \\
\hline
(4,1)    & 1.069  & 0.461  & 1.219  \\
(4,3/2)  & 0.575  & 0.957  & 0.316  \\
(4,2)    & 0.0822 & 1.453  & 0.0298 \\
\hline
(5,1/2)  & 1.397  & 0.0964 & 7.630  \\
(5,1)    & 0.904  & 0.592  & 0.804  \\
(5,3/2)  & 0.411  & 1.088  & 0.199  \\
\hline
(6,1/2)  & 1.233  & 0.227  & 2.856  \\
(6,1)    & 0.740  & 0.723  & 0.539  \\
(6,3/2)  & 0.247  & 1.219  & 0.106  \\
\hline
(7,1/2)  & 1.069  & 0.358  & 1.571  \\
(7,1)    & 0.575  & 0.854  & 0.355  \\
(7,3/2)  & 0.0822 & 1.349  & 0.0321 \\
\hline
(8,1/2)  & 0.904  & 0.489  & 0.973  \\
(8,1)    & 0.411  & 0.985  & 0.220  \\
\hline
(9,0)    & 1.233  & 0.124  & 5.236  \\
(9,1/2)  & 0.740  & 0.620  & 0.628  \\
(9,1)    & 0.247  & 1.115  & 0.116  \\
\hline
(10,0)   & 1.069  & 0.255  & 2.208  \\
(10,1/2) & 0.575  & 0.750  & 0.4035 \\
(10,1)   & 0.0822 & 1.246  & 0.0347 \\
\hline
(11,0)   & 0.904  & 0.386  & 1.234  \\
(11,1/2) & 0.411  & 0.881  & 0.245  \\
\hline
(12,0)   & 0.740  & 0.516  & 0.754  \\
(12,1/2) & 0.247  & 1.012  & 0.128  \\
\hline
(13,0)   & 0.575  & 0.647  & 0.468  \\
(13,1/2) & 0.0822 & 1.143  & 0.03785 \\
\hline
(14,0)   & 0.411  & 0.778  & 0.278  \\
\hline
(15,0)   & 0.247  & 0.909  & 0.143  \\
\hline
(16,0)   & 0.0822 & 1.040  & 0.0416 \\
\hline\hline
\end{tabular}
\end{center}
\label{fadj_su3_theories_table}
\end{table}


\begin{table}
  \caption{\footnotesize{Values of the coefficients $\kappa^{(F)}_j$, 
$j=1,2,3$, for the scheme-independent expansion of the anomalous dimension 
$\gamma_{\bar\psi\psi,IR}$ in an SU(3) gauge theory with fermions in the 
fundamental and adjoint representations, as functions of $N_{Adj}$. 
Half-integral values of $N_{Adj}$ correspond to $2N_{Adj}$
copies of Majorana fermions in the adjoint representation. The notation 
$a$e-n means $a \times 10^{-n}$.}}
\begin{center}
\begin{tabular}{|c|c|c|c|} \hline\hline
$N_{Adj}$ & $\kappa^{(F)}_1$ & $\kappa^{(F)}_2$ & $\kappa^{(F)}_3$ 
\\ \hline
0   & 4.98e-2 & 3.79e-3 & 2.37e-4 \\
$\frac{1}{2}$ 
    & 4.56e-2 & 3.39e-3 & 1.835e-4 \\
1   & 4.20e-2 & 3.03e-3 & 1.51e-4 \\
$\frac{3}{2}$
    & 3.89e-2 & 2.71e-3 & 1.31e-4 \\
2   & 3.63e-2 & 2.44e-3 & 1.16e-4 \\
\hline\hline
\end{tabular}
\end{center}
\label{kappa_fund_su3fa_table}
\end{table}


\begin{table}
  \caption{\footnotesize{Values of the coefficients $\kappa^{(Adj)}_j$,
$j=1,2,3$, for the scheme-independent expansion of the anomalous dimension
$\gamma_{\bar\chi\chi,IR}$ in an SU(3) gauge theory with fermions in the
fundamental and adjoint representations, for illustrative values of $N_F$.}}
\begin{center}
\begin{tabular}{|c|c|c|c|} \hline\hline
$N_F$ & $\kappa^{(Adj)}_1$ & $\kappa^{(Adj)}_2$ & $\kappa^{(Adj)}_3$
\\ \hline
0   & 0.444 & 0.234 & 0.121 \\
4   & 0.484 & 0.270 & 0.145 \\
8   & 0.532 & 0.315 & 0.179 \\
10  & 0.560 & 0.342 & 0.201 \\
12  & 0.590 & 0.372 & 0.227 \\
\hline\hline
\end{tabular}
\end{center}
\label{kappa_adj_su3fa_table}
\end{table}


\begin{table}
  \caption{\footnotesize{Values of $N_{F,u}$ from Eq. (\ref{Nfu}) and
      $N_{Adj,u}$ from Eq. (\ref{Nf2u}) (formally generalized to non-negative
      real numbers) for the illustrative SU(3) 
theories with $N_F$ fermions in the fundamental representation and $N_{Adj}$
fermions in the adjoint representation.  Half-integral values of $N_{Adj}$
refer to theories with $2N_{Adj}$ Majorana fermions in the adjoint
representation.}}
\begin{center}
\begin{tabular}{|c|c|c|} \hline\hline
$(N_F,N_{Adj})$    & $N_{F,u}$         &  $N_{Adj,u}$      \\ 
\hline
(8,1/2)  & 27/2  & 17/12  \\
(8,1)    & 21/2  & 17/12  \\
\hline
(10,0)    & 33/2 & 13/12  \\
(10,1/2)  & 27/2 & 13/12  \\
(10,1)    & 21/2 & 13/12  \\
\hline
(12,0)    & 33/2 & 3/4    \\
(12,1/2)  & 27/2 & 3/4    \\
\hline\hline
\end{tabular}
\end{center}
\label{nup_values_su3_fa}
\end{table}


%
\begin{table}
\caption{\footnotesize{Values of the anomalous dimension 
 $\gamma_{\bar\psi\psi,IR,\Delta_F^p}$, 
 calculated to order $p=1, \ 2, \ 3$ and evaluated at the IR fixed point 
in an SU(3) gauge theory with 
$N_F$ fermions in the fundamental ($F$) representation and $N_{Adj}$ fermions
in the adjoint ($Adj$) representation. Here, $\psi$ is the fermion in the $F$
representation. }}
\begin{center}
\begin{tabular}{|c|c|c|c|c|c|c|} \hline\hline
$(N_F,N_{Adj})$   & $\gamma_{\bar\psi\psi,IR,\Delta_F}$ & 
                    $\gamma_{\bar\psi\psi,IR,\Delta_F^2}$ & 
                    $\gamma_{\bar\psi\psi,IR,\Delta_F^3}$
\\ \hline
(8,1/2)  & 0.251  & 0.353  & 0.384  \\
(8,1)    & 0.105  & 0.124  & 0.126  \\
\hline
(10,0)   & 0.324  & 0.484  & 0.549  \\
(10,1/2) & 0.159  & 0.201  & 0.209  \\
(10,1)   & 0.0210 & 0.0218 & 0.0218 \\
\hline
(12,0)   & 0.224  & 0.301  & 0.323  \\
(12,1/2) & 0.0684 & 0.0760 & 0.0766 \\
\hline\hline
\end{tabular}
\end{center}
\label{gamma_fund_fa_su3_values}
\end{table}


%
\begin{table}
\caption{\footnotesize{Values of the anomalous dimension
 $\gamma_{\bar\chi\chi,IR,\Delta_{Adj}^p}$,
 calculated to order $p=1, \ 2, \ 3$ and evaluated at the IR fixed point
in an SU(3) gauge theory with
$N_F$ fermions in the fundamental ($F$) representation and $N_{Adj}$ fermions
in the adjoint ($Adj$) representation. Here, $\chi$ is the fermion in the
$Adj$ representation.}}
\begin{center}
\begin{tabular}{|c|c|c|c|c|c|c|} \hline\hline
$(N_F,N_{Adj})$    & $\gamma_{\bar\chi\chi,IR,\Delta_{Adj}}$ &
                    $\gamma_{\bar\chi\chi,IR,\Delta_{Adj}^2}$ &
                    $\gamma_{\bar\chi\chi,IR,\Delta_{Adj}^3}$
\\ \hline
(8,1/2)   & 0.488  & 0.753  & 0.891  \\
(8,1)     & 0.222  & 0.276  & 0.289  \\
\hline
(10,1/2)  & 0.326  & 0.443  & 0.483  \\
(10,1)    & 0.0466 & 0.0490 & 0.0491 \\
\hline
(12,1/2)  & 0.1475 & 0.171  & 0.174  \\
\hline\hline
\end{tabular}
\end{center}
\label{gamma_adj_fa_su3_values}
\end{table}


\begin{table}
  \caption{\footnotesize{Values of the coefficients $d_j$, 
$j=2,3,4$, for the scheme-independent expansion of $\beta'_{IR}$,
Eq. (\ref{betaprime_ir_Deltaseries_fa}), 
in an SU(3) gauge theory with fermions in the 
fundamental and adjoint representations, as functions of $N_{Adj}$. 
Half-integral values of $N_{Adj}$ correspond to $2N_{Adj}$
copies of Majorana fermions in the adjoint representation. The notation 
$a$e-n means $a \times 10^{-n}$.}}
\begin{center}
\begin{tabular}{|c|c|c|c|} \hline\hline
$N_{Adj}$ & $d_2$ & $d_3$ & $d_4$ 
\\ \hline
0   & 0.831e-2 & 0.983e-3  & $-$0.463e-4  \\
1/2 & 0.760e-2 & 0.8225e-3 & $-$2.44e-5   \\
1   & 0.700e-2 & 0.698e-3  & $-$1.24e-5   \\
3/2 & 0.649e-2 & 0.600e-3  & $-$0.578e-5  \\
2   & 0.605e-2 & 0.521e-3  & $-$2.12e-6   \\
\hline\hline
\end{tabular}
\end{center}
\label{d_fund_su3fa_table}
\end{table}


\begin{table}
  \caption{\footnotesize{Values of the coefficients $\tilde d_j$, 
$j=2,3,4$, for the scheme-independent expansion of $\beta'_{IR}$, Eq. 
(\ref{betaprime_ir_Deltaseries2_fa}), 
in an SU(3) gauge theory with fermions in the 
fundamental and adjoint representations, for illustrative values of $N_F$.}}
\begin{center}
\begin{tabular}{|c|c|c|c|} \hline\hline
$N_F$ & $\tilde d_2$ & $\tilde d_3$ & $\tilde d_4$ 
\\ \hline
0   & 0.1975  & 0.117   & 0.0265   \\
4   & 0.215   & 0.139   & 0.0313   \\
8   & 0.236   & 0.168   & 0.0358   \\
10  & 0.249   & 0.186   & 0.0374   \\
12  & 0.262   & 0.206   & 0.0379   \\
\hline\hline
\end{tabular}
\end{center}
\label{dtilde_su3fa_table}
\end{table}


%
\begin{table}
\caption{\footnotesize{Values of $\beta'_{IR}$ as calculated to order 
$O(\Delta_f^p)$ via Eq. (\ref{betaprime_ir_Deltaseries_fa}), 
denoted $\beta'_{IR,\Delta_F^p}$ and 
to order $O(\Delta_{Adj}^p)$ via Eq. (\ref{betaprime_ir_Deltaseries2_fa}),
denoted $\beta'_{IR,\Delta_{Adj}^p}$, with $p=2,3,4$, 
in an SU(3) gauge theory with $N_F$
fermions in the fundamental ($F$) representation and $N_{Adj}$ fermions in the
adjoint ($Adj$) representation. Here, half-integral values of $N_{Adj}$ refer to
theories with $2N_{Adj}$ copies of Majorana fermions in the adjoint
representation. The notation $a$e-n means $a \times 10^{-n}$.}}
\begin{center}
\begin{tabular}{|c|c|c|c|c|c|c|} \hline\hline
$(N_F,N_{Adj})$    & $\beta'_{IR,\Delta_F^2}$ & $\beta'_{IR,\Delta_{Adj}^2}$ & 
                     $\beta'_{IR,\Delta_F^3}$ & $\beta'_{IR,\Delta_{Adj}^3}$ &
                     $\beta'_{IR,\Delta_F^4}$ & $\beta'_{IR,\Delta_{Adj}^4}$ 
\\ \hline
(8,1/2)  & 0.230   & 0.199   & 0.367 & 0.328  & 0.344 & 0.353  \\
(8,1)    & 4.374e-2& 4.105e-2& 5.465e-2&5.32e-2&5.42e-2 & 5.43e-2 \\
\hline
(10,0)   & 0.351   & 0.292   & 0.621 & 0.528  & 0.538 & 0.579  \\
(10,1/2) & 0.0931  & 0.0846  & 0.128 & 0.1215 & 0.125 & 0.126  \\
(10,1)   & 1.75e-3  & 1.73e-3 & 1.837e-3&1.8345e-3&1.8363e-3&1.8361e-3 \\
\hline
(12,0)   & 0.168   & 0.1475  & 0.258 & 0.235  & 0.239 & 0.247     \\
(12,1/2) & 1.71e-2&1.64e-2&1.987e-2& 1.962e-2&1.975e-2&1.977e-2 \\
\hline\hline
\end{tabular}
\end{center}
\label{betaprime_fa_su3_values}
\end{table}
%



\begin{thebibliography}{99}

\bibitem{fm}
Taking the fermions to be massless does not incur any loss of generality, 
because a fermion with a nonzero mass $m_0$ would be
integrated out of the low-energy effective field theory at Euclidean momentum 
scales $\mu < m_0$, and hence would be irrelevant to the properties of the
theory at the IRFP of interest here.  We note also that our theories do not 
contain any scalar fields. 

\bibitem{rs09}
T. A. Ryttov and F. Sannino, Int. J. Mod. Phys. {\bf 25}, 4603 (2010).

\bibitem{bv}
T. A. Ryttov and R. Shrock, Phys. Rev. D {\bf 81}, 116003 (2010) 

\bibitem{scalecon}
 J. Polchinski, Nucl. Phys. B {\bf 303}, 226 (1988);
 J.-F. Fortin, B. Grinstein and A. Stergiou, JHEP 01 (2013) 184 (2013);
 A. Dymarsky, Z. Komargodski, A. Schwimmer, and S. Thiessen, JHEP {\bf 10},
 171 (2015) and references therein.

\bibitem{bvh} 
T. A. Ryttov, R. Shrock, Phys. Rev. D {\bf 83}, 056011 (2011)
[arXiv:1011.4542]. 

\bibitem{ps}
C. Pica, F. Sannino, Phys. Rev. D {\bf 83}, 035013 (2011) [arXiv:1011.5917].

\bibitem{bc}
R. Shrock, Phys. Rev. D {\bf 87}, 105005 (2013) [arXiv:1301.3209]; 
Phys. Rev. D {\bf 87}, 116007 (2013) [arXiv:1302.5434].

\bibitem{flir}
T. A. Ryttov and R. Shrock, Phys. Rev. D {\bf 94}, 105015 (2016)
[arXiv:1607.06866]. 

\bibitem{bz}
T. Banks and A. Zaks, Nucl. Phys. B {\bf 196}, 189 (1982).

\bibitem{gkgg}
G. Grunberg, Phys. Rev. D {\bf 46}, 2228 (1992);
E. Gardi and M. Karliner, Nucl. Phys. B {\bf 529}, 383 (1998);
E. Gardi and G. Grunberg, JHEP 03, 024 (1999).

\bibitem{gtr}
T. A. Ryttov, Phys. Rev. Lett. {\bf 117}, 071601 (2016) [arXiv:1604.00687].

\bibitem{gsi}
T. A. Ryttov and R. Shrock, Phys. Rev. D {\bf 94}, 105014 (2016)
[arXiv:1608.00068].

\bibitem{dex}
T. A. Ryttov and R. Shrock, Phys. Rev. D {\bf 94}, 125005 (2016) 
[arXiv:1610.00387].

\bibitem{dexs}
T. A. Ryttov  and R. Shrock, Phys. Rev. D {\bf 95}, 085012 (2017)
[arXiv:1701.06083].

\bibitem{dexl}
T. A. Ryttov  and R. Shrock, Phys. Rev. D {\bf 95}, 105004 (2017) 
[arXiv:1703.08558].

\bibitem{dexo}
T. A. Ryttov  and R. Shrock, Phys. Rev. D {\bf 96}, 105015 (2017) 
[arXiv:1709.05358]. 

\bibitem{pgb}
T. A. Ryttov  and R. Shrock, Phys. Rev. D {\bf 97}, 025004 (2018) 
[arXiv:1710.06944].

\bibitem{dexss}
T. A. Ryttov  and R. Shrock, Phys. Rev. D {\bf 96}, 105018 (2017)
[arXiv:1706.06422]; Phys. Rev. D {\bf 97}, 065020 (2018) [arXiv:1711.01116].

\bibitem{bpcgt}
T. A. Ryttov  and R. Shrock, Phys. Rev. D {\bf 97}, 016020 (2018)
[arXiv:1710.00096]. 

\bibitem{lgtreviews}
%
For recent reviews of these lattice simulations, see, 
e.g., talks in the Lattice for BSM 2017 Workshop at
http://www-hep.colorado.edu/(tilde)eneil/lbsm17;
Lattice-2017 at http://wpd.ugr.es/(tilde)lattice2017; \cite{simons}; and
Lattice-2018 at https://web.pa.msu.edu/conf/Lattice2018.
At present, there is not a complete consensus among lattice groups about the
respective values of $N_{f,cr}$ for various theories. 

\bibitem{simons}
Simons Workshop on Continuum and Lattice Approaches to the Infrared Behavior of
Conformal and Quasiconformal Gauge Theories, Jan. 8-12, 2018, T. A. Ryttov and
R. Shrock, organizers; http://scgp.stonybrook.edu/archives/21358.

\bibitem{lnncomment}
%
Furthermore, in the case where
$G={\rm SU}(N_c)$ and the fermions are in the fundamental representation,
one can take the limits $N_c \to \infty$ and
$N_f \to \infty$ with $r=N_f/N_c$ fixed and finite.  In this case,
$\alpha_{IR}$ can be made arbitrarily small. 

\bibitem{gammaconvention}
%
Some authors use the opposite sign convention for the anomalous dimension,
writing $D_{\cal O} = D_{\cal O,{\rm free}} + \gamma_{\cal O}$. Our sign
convention is the same as the convention used in lattice gauge
theory literature.

\bibitem{gracey_gammatensor}
J. A. Gracey, Phys. Lett. B {\bf 488}, 175 (2000).

\bibitem{ms}
G. 't Hooft, Nucl. Phys. {\bf B61}, 455 (1973).

\bibitem{msbar}
W. A. Bardeen, A. J. Buras, D. W. Duke, and T. Muta, Phys. Rev. D {\bf 18},
3998 (1978).

\bibitem{gross75}
D. J. Gross, in R. Balian and J. Zinn-Justin, eds.
{\it Methods in Field Theory}, Les Houches 1975
(North Holland, Amsterdam, 1976), p. 141.

\bibitem{b1}
D. J. Gross and F. Wilczek, Phys. Rev. Lett. {\bf 30}, 1343 (1973);
H. D. Politzer, Phys. Rev. Lett. {\bf 30}, 1346 (1973). 

\bibitem{b2}
W. E. Caswell, Phys. Rev. Lett. {\bf 33}, 244 (1974);
D. R. T. Jones, Nucl. Phys. B {\bf 75}, 531 (1974).

\bibitem{b3}
O. V. Tarasov, A. A. Vladimirov, and A. Yu. Zharkov, Phys. Lett. B {\bf 93},
429 (1980); S. A. Larin and J. A. M. Vermaseren, Phys. Lett. B {\bf 303}, 334
(1993).

\bibitem{b4}
T. van Ritbergen, J. A. M. Vermaseren, and S. A. Larin, Phys. Lett. B 
{\bf 400}, 379 (1997). 

\bibitem{b5}
F. Herzog, B. Ruijl, T. Ueda, J. A. M. Vermaseren, and A. Vogt,
JHEP 02 (2017) 090. 

\bibitem{b5su3}
P. A. Baikov, K. G. Chetyrkin, and J. H. K\"uhn, 
Phys. Rev. Lett. {\bf 118}, 082002 (2017). 

\bibitem{zoller}
M. F. Zoller, JHEP 10, 118 (2016) [arXiv:1608.08982]. 

\bibitem{c4}
K. G. Chetyrkin, Phys. Lett. B {\bf 404}, 161 (1997);
J. A. M. Vermaseren, S. A. Larin, and T. van Ritbergen, Phys. Lett. B 
{\bf 405}, 327 (1997). 

\bibitem{c5}
P. A. Baikov, K. G. Chetyrkin, and J. H. K\"uhn, JHEP 10, 076 (2014);
JHEP 04 (2017) 119.

\bibitem{chetzol}
K. G. Chetyrkin and M. F. Zoller, JHEP 06, 074 (2017) [arXiv:1704.04209].

\bibitem{cgs}
J. A. Gracey, T. A. Ryttov, and R. Shrock, Phys. Rev. D {\bf 97}, 116018 
(2018) [arXiv:1805.02729].

\bibitem{phasenote}
%
In principle, for certain $G$, $R$, and $R'$, it might be possible for there to
be an intermediate phase between the (deconfined) non-Abelian Coulomb phase and
the QCD-like phase with confinement and S$\chi$SB with the property that in
this intermediate phase there is confinement but no S$\chi$SB. A necessary but
not sufficient condition for this would be that the 't Hooft anomaly-matching 
conditions are satisfied.  This possibility will not be directly relevant for
our calculations in the NACP and so we will not pursue it here. 

\bibitem{su4lgt1}
V. Ayyar, T. DeGrand, M. Golterman, D. Hackett, W. I. Jay, E. T. Neil, 
Y. Shamir, and B. Svetitsky, Phys. Rev. D {\bf 97}, 074505 (2016) 
[arXiv:1710.00806].

\bibitem{su4lgt2}
V. Ayyar, T. DeGrand, D. Hackett, W. I. Jay, E. T. Neil, 
Y. Shamir, and B. Svetitsky, Phys. Rev. D {\bf 97}, 114505 (2018) 
[arXiv:1801.05809]; Phys. Rev. D {\bf 97}, 114502 (2018) [arXiv:1802.09644].

\bibitem{traceanomaly}
S. L. Adler, J. C. Collins, and A. Duncan, Phys. Rev. D {\bf 15}, 1712 (1977);
J. C. Collins, A. Duncan, and S. Joglekar, Phys. Rev. D {\bf 16}, 438
(1977); N. K. Nielsen, Nucl. Phys. B {\bf 120}, 212 (1977); see also
H. Kluberg-Stern and J.-B. Zuber, Phys. Rev. D {\bf 12}, 467 (1975).

\bibitem{tracerel}
See, e.g., S. S. Gubser, A. Nellore, S. S. Pufu, and E. D. Rocha,
Phys. Rev. Lett. {\bf 101}, 131601 (2008); see also M. Kurachi, S. Matsuzaki, 
and K. Yamawaki, Phys. Rev. D {\bf 90}, 055028 (2014); 
R. J. Crewther and L. C. Tunstall, Phys. Rev. D {\bf 91}, 034016 (2015). 

\bibitem{rsv}
T. van Ritbergen, A. N. Schellekens, and J. A. M. Vermaseren,
Int. J. Mod. Phys. A {\bf 14}, 41 (1999).

\end{thebibliography}
\end{document}